\documentclass[]{tCPH2e}

\newcommand{\SIGMA}{\mbox{\boldmath${\sigma}$}}

\newcommand{\NABLA}{\mbox{\boldmath${\nabla}$}}
\newcommand{\GAMMA}{\mbox{\boldmath${\gamma}$}}

\newcommand{\PARTIAL}{\mbox{\boldmath${\partial}$}}
\newcommand{\OMEGA}{\mbox{\boldmath${\Omega}$}}
\newcommand{\be}{\begin{equation}}
\newcommand{\ee}{\end{equation}}
\newcommand{\bq}{\begin{eqnarray}}
\newcommand{\eq}{\end{eqnarray}}

\newcommand{\p}{{\bf{p}}}

\newcommand{\kk}{{\bf k}}
\newcommand{\ii}{{\bf i}}
\newcommand{\jj}{{\bf j}}
\newcommand{\rr}{{\bf r}}

\newcommand{\ket}[1]{\left | \, #1 \right\rangle}

\begin{document}
\doi{10.1080/0010751YYxxxxxxxx}
 \issn{1366-5812}
\issnp{0010-7514}

\jvol{00} \jnum{00} \jyear{2008} \jmonth{June}

\markboth{Taylor \& Francis and I.T. Consultant}{Contemporary Physics}

\articletype{RESEARCH ARTICLE}

\title{\itshape Manifestations of topological effects in graphene}

\author{Jiannis K. Pachos$^a$$^{\ast}$\thanks{$^\ast$Corresponding author. Email:
j.k.pachos@leeds.ac.uk, URL: quantum.leeds.ac.uk/$\sim$phyjkp
$^{a}${\em{School of Physics \& Astronomy, University of Leeds, Leeds LS2
9JT, UK}}}}

\maketitle

\begin{abstract}

Graphene is a monoatomic layer of graphite with Carbon atoms arranged in a
two dimensional honeycomb lattice configuration. It has been known for more
than sixty years that the electronic structure of graphene can be modelled
by two-dimensional massless relativistic fermions. This property gives rise
to numerous applications, both in applied sciences and in theoretical
physics. Electronic circuits made out of graphene could take advantage of
its high electron mobility that is witnessed even at room temperature. In
the theoretical domain the Dirac-like behavior of graphene can simulate high
energy effects, such as the relativistic Klein paradox. Even more
surprisingly, topological effects can be encoded in graphene such as the
generation of vortices, charge fractionalization and the emergence of
anyons. The impact of the topological effects on graphene's electronic
properties can be elegantly described by the Atiyah-Singer index theorem.
Here we present a pedagogical encounter of this theorem and review its
various applications to graphene. A direct consequence of the index theorem
is charge fractionalization that is usually known from the fractional
quantum Hall effect. The charge fractionalization gives rise to the exciting
possibility of realizing graphene based anyons that unlike bosons or
fermions exhibit fractional statistics. Besides being of theoretical
interest, anyons are a strong candidate for performing error free quantum
information processing.

\bigskip \bigskip

\end{abstract}

\section{Introduction}

It is a rare example in science to have theoretical developments and
technological applications growing simultaneously at the same rate as with
research in graphene~\cite{Geim,Neto}. This two dimensional monatomic layer
of graphite was theoretically first investigated by Wallace~\cite{Wallace}.
Experimentally it was successfully isolated at the University of
Manchester~\cite{Novoselov04}. Graphene consists of carbon atoms positioned
at the vertices of a two dimensional honeycomb lattice. Its electronic
properties can be modelled by free electrons that are allowed to tunnel from
one site of the honeycomb lattice to a neighboring one. The simple
geometrical configuration of the Carbon atoms gives graphene, in the
microscopic level, a surprisingly rich collective behavior. The latter is
witnessed by quantum effects so resilient that they can influence the
behavior of graphene even at room temperature. Due to its unique character
graphene can successfully simulate phenomena that are expected to appear in
relativistic quantum theories, such as the Klein paradox~\cite{Geim_klein},
and it revealed new effects, such is the anomalous quantum Hall
effect~\cite{Novoselov05,Hall}.

The desire to employ graphene for studying topological effects comes from
the simplicity with which these effects are encoded in graphene. The
Dirac-like description of graphene gives a natural starting point. By
varying the geometry of graphene one can create topologically distinct
configurations such as a sphere or a torus on which the effective Dirac
equation is defined. Moreover, by perturbing the graphene lattice and the
tunnelling couplings of the electrons one can give rise to effective gauge
and scalar fields. These fields are naturally coupled to the Dirac fermions
that model the behavior of graphene. One can encode vortices on the
effective scalar field with vorticity determined by the flux of the gauge
field going through the vortex, much like in superconductors. How the Dirac
fermions respond to these topological defects is the central subject of this
article.

\subsection{The structure of graphene}

\begin{figure}
\begin{center}
\includegraphics[width=8cm]{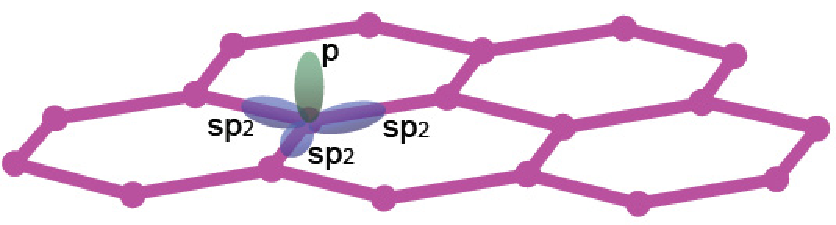}
\caption{The graphene molecule with a $C$ atom placed at each site
of a honeycomb lattice. The electron orbitals of each $C$ atom are depicted.
Hybridization provides three equivalent $sp_2$ orbitals that create the
strong $\sigma$ bonds with the neighboring atoms. This gives rise to the
covalent ``backbone" of graphene with honeycomb lattice geometry. The
remaining $2p_z$ orbital creates a half filled $\pi$ bond. The latter
provides free electrons that tunnel from site to site giving rise to the
fascinating electronic properties of graphene.}
\label{fig:bonds}
\end{center}
\end{figure}

Graphene is a large molecule of carbon atoms, $C$, that are strongly bound
together in a flat configuration. The $C$ atoms are positioned on the sites
of a honeycomb lattice, as shown in Fig.~\ref{fig:bonds}. Three of the four
outer electrons of each carbon atom are employed to strongly bond with its
three neighboring atoms via $\sigma$ bonds. The $2p_z$ orbital of the fourth
electron creates a $\pi$ bond with a neighboring $C$ atoms. The $\sigma$
bonds provide the covalent ``backbone" structure of graphene with honeycomb
lattice geometry. Their strength is responsible for the flexibility and
robustness of the lattice. The $\pi$ bonds give rise to graphene's unique
electronic structure. Each $\pi$ bond is half filled allowing the $p$
orbital electrons to tunnel from one atom to the neighboring one. Graphene
should be considered as a many body system: several electrons are allowed to
tunnel from site to site, while at the same time they satisfy the Pauli
exclusion principle. In the following we use this simple model to study the
properties of graphene.

\subsection{Topological properties}

\begin{figure}[h]
\begin{center}
\includegraphics[width=12cm]{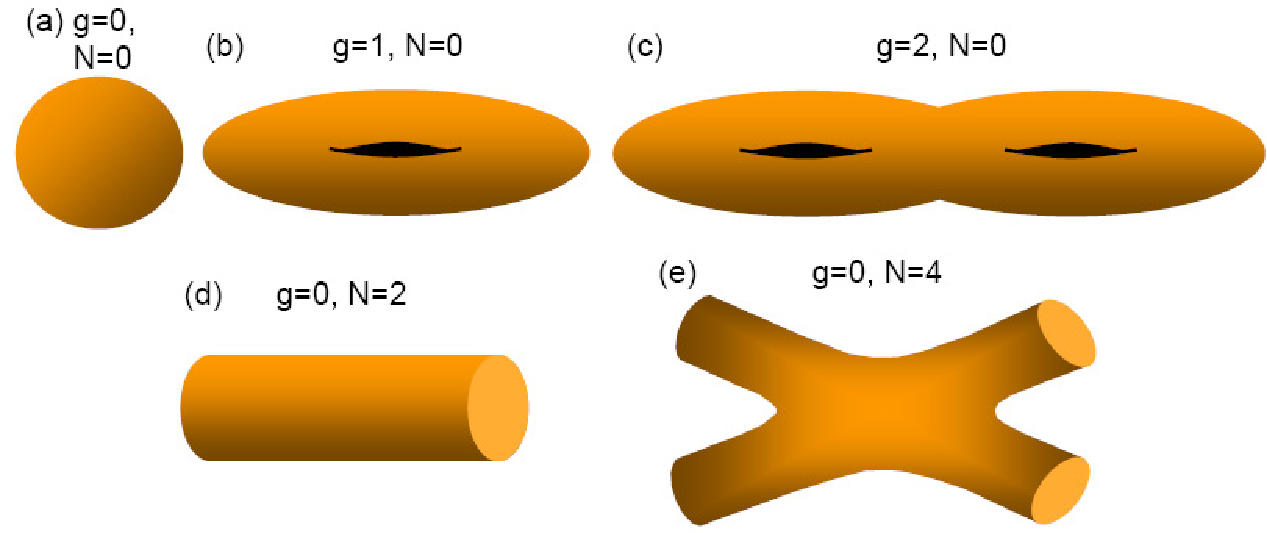}
\caption{Two dimensional surfaces with different topologies characterized
by the genus, $g$, or the number of open faces, $N$. The surfaces (a), (b)
and (c) are compact with $N=0$ and $g=0$, $1$ and $2$, respectively. The
surfaces (e) and (d) are open with $g=0$ and $N=2$ and $4$, respectively.}
\label{fig:topologies}
\end{center}
\end{figure}

Topology is a branch of mathematics that is concerned with the global
properties of geometrical objects rather than with their local details.
These properties are supposed to be unaffected by continuous deformations of
the objects. Restricted in two dimensions we can consider a surface that can
be compact, like a sphere or a torus, or it can be open, like a flat sheet
or a cylinder. These characteristics are very general and they are invariant
even if small geometrical deformations are caused on the surface such as
ripples. Compact surfaces can be topologically equivalent to a sphere, or to
a torus, if they have a hole, like a ring doughnut. More complicated
topologies are possible, too. To distinguish them we assign an integer
number called the genus, $g$, that gives the number of ``holes" of each
surface. Examples of compact surfaces with genus $g=0$, $g=1$ and $g=2$ are
seen in Fig.~\ref{fig:topologies}(a), (b) and (c), respectively. It is clear
that $g$ is invariant even under deformations of the surface geometry as
long as no holes are created or destroyed. In that spirit a cup is
topologically equivalent to a ring!

It is a natural question to ask how graphene, a naturally appearing
material, can possess such extraordinary properties as the ones related to
topology. The main reason is that its chemical bonding is so strong that it
induces a stability in the geometrical properties of the underlined lattice.
Indeed, several stable geometric variants of graphene appear in nature. In
all of them each carbon atom has precisely three neighbors. The lattice of
these molecules mainly consists of hexagons with a number of additional
pentagons or heptagons that induce curvature. For example, one can encounter
cylinders (Fig.~\ref{fig:topologies}(d)), so called nanotubes, produced from
a folded graphene sheet, with the number of open faces, $N$, being equal to
two. Spherical configurations are also possible
(Fig.~\ref{fig:topologies}(a)) the so called fullerenes with $N=0$. The same
geometry is also met in the football, where twelve pentagons are present in
between the hexagonal plaquettes so that a spherical configuration can be
created. It is also possible to have crossing nanotubes corresponding to a
surface with four open faces, as seen in Fig.~\ref{fig:topologies}(e). Apart
from the robust geometrical structure of the underlined lattice, the
presence of long range quantum coherences are also responsible for the
emergence of the topological character in graphene. Indeed, at low enough
temperatures one can assume that graphene's electrons can travel coherently
through the whole molecule, thus detecting its geometry or topology. This
makes graphene the natural laboratory to study the interplay between
topology and quantum physics.

\subsection{The index theorem}

Classical as they may be, topological configurations can have a dramatic
effect on the quantum behavior of a system. In particular, they determine
the possible quantum states a system can have. This striking connection is
demonstrated by the index theorem, initially introduced by Atiyah and Singer
in the sixties~\cite{Atiyah}. The index theorem provides a direct way to
determine certain properties of the spectrum of general Hamiltonians. Its
application is rather straightforward providing a good alternative to the
direct diagonalization of the Hamiltonian. This can be a tantalizing task
for large systems. It is, therefore, not hard to imagine that the index
theorem can have a dramatic impact on theoretical and applied
sciences~\cite{Eguchi,Stone}.

\begin{figure}[h]
\begin{center}
\includegraphics[width=3cm]{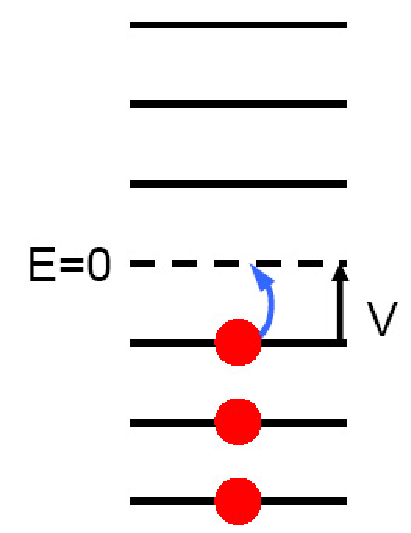}
\caption{The energy levels of a fermionic system are considered here to be
symmetric around $E=0$. For half filling systems there are half as many
fermions as there are possible states. The lowest energy is obtained by
occupying the negative energy states with a single fermion due to the Pauli
exclusion principle. The application of a properly tuned potential
difference, $V$, can cause the system to change its state, if there are zero
modes to be occupied. This will give rise to an electric current. In the
absence of zero modes no change will be observed and the system will appear
as an insulator.}
\label{fig:Fermi}
\end{center}
\end{figure}

In particular, the index theorem provides information about the number of
zero energy states that are present in systems such as graphene and its
geometric variants~\cite{Pachos,Pachos1}. They are important for determining
conductivity properties. To understand why let us look closer at the nature
of these molecules. It is common that the employed Hamiltonians have
symmetric spectra around zero energy, called the particle-hole symmetry. The
negative part of the spectra is called valence band and the positive part
conducting band. As mentioned above, graphene has approximately half as many
free electrons as the positions they can occupy. Due to the Pauli exclusion
principle these electrons fill up the valence band states one by one until
the zero energy point. This point corresponds to half of the possible
occupation positions. When adding a small potential difference one can
witness conducting or insulating behavior depending on the presence of zero
modes. If there are zero modes then they can be occupied in response to the
electric field. This change in the electronic states appears macroscopically
as a current. If there are no zero modes the small shift in the energy will
cause no change in the behavior of the system, thus appearing as an
insulator.

Apart from the conducting properties, the presence of zero modes can give
rise to exotic characteristics. It can be shown by general arguments that
vortices encoded on graphene can cause the fractionalization of charge. This
collective behavior is known from the fractional quantum Hall effect.
Drawing the analogy further, the vortices in graphene could actually be
possessing fractional anyonic statistics, which is neither similar to bosons
nor fermions~\cite{Brennen}. It is an exciting possibility to have a
physical system that needs mainly isolation from the environment to reveal
such fascinating properties. Experiments that reveal such aspects of
topological behavior are usually hard to perform due to the complexity
needed for topological properties to arise. Graphene, however, could achieve
this with minimal requirements.

\section{Elementary properties of graphene}

\begin{figure}[h]
\begin{center}
\includegraphics[width=3.5cm]{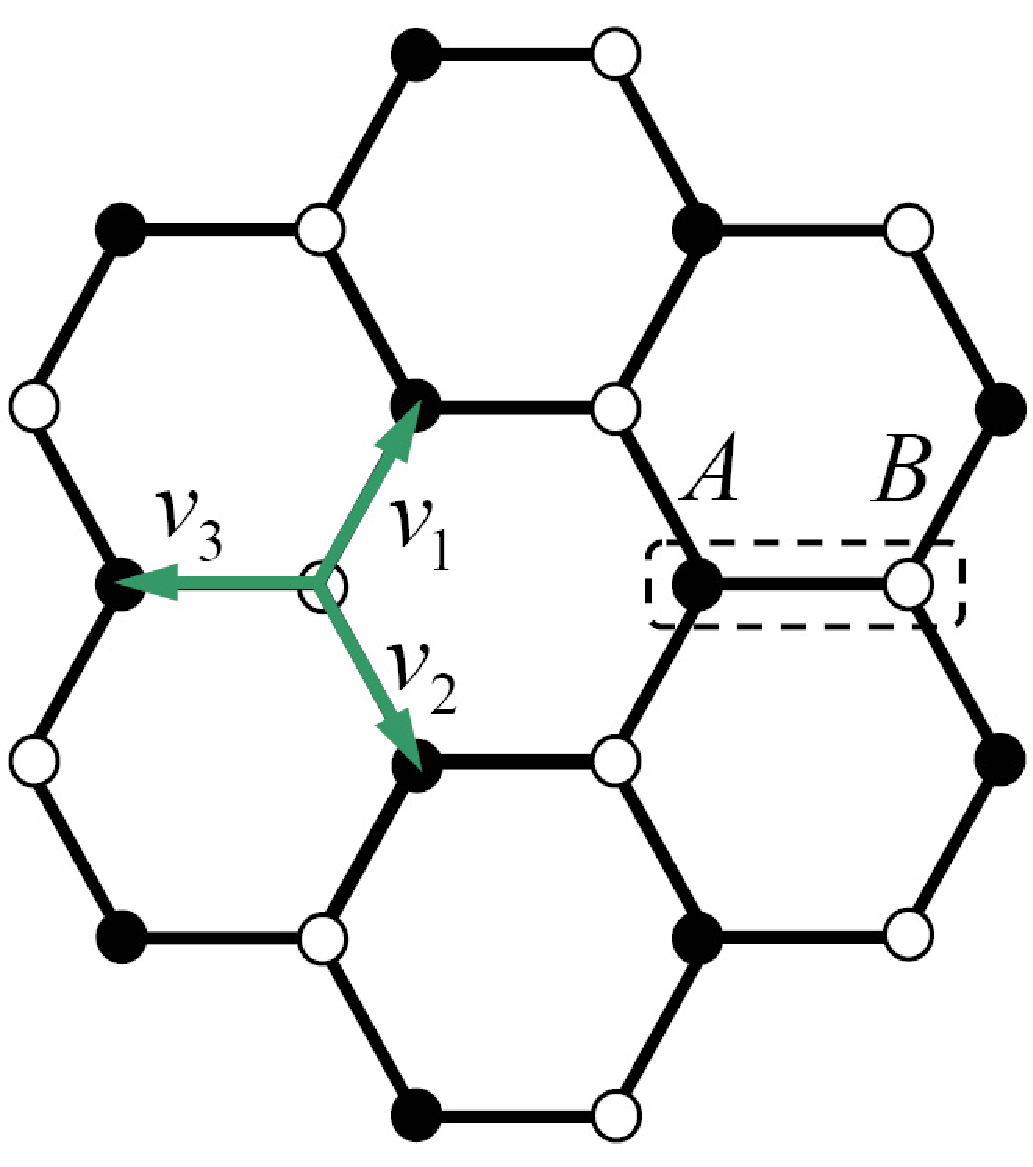}
\caption{Graphene can be modelled by a honeycomb lattice, where electrons are
tunnelling from site to site. This lattice has hexagons as the elementary
plaquettes. It can be split into two triangular sublattices, indicated by
black and white circles and denoted here by $A$ and $B$, respectively. The
doted rectangular denotes the unit cell of the lattice.}
\label{fig:honeycomb}
\end{center}
\end{figure}

As we have seen, graphene is a molecule consisting of $C$ atoms arranged on
a two dimensional honeycomb lattice. The elementary plaquette of the lattice
is a hexagon and the atoms are placed on the sites of the lattice.
Graphene's electronic properties can efficiently be described by the tight
binding model, where spinless electrons tunnel from site to site along the
links of the lattice without interacting with each other. This simple model
provides a surprisingly accurate description of graphene and gives rise to a
wealth of phenomena. The tight binding Hamiltonian is given by
\begin{equation}
H = -J\sum_{\langle \ii,\jj\rangle} (a^\dagger_{\ii} a_{\jj}+a^\dagger_{\jj}
a_{\ii}),
\label{Ham1}
\end{equation}
where the operator $a_{\ii}^\dagger$ creates an electron at site $\ii$,
$a_{\jj}$ annihilates it at $\jj$ and the coupling $J$ gives the strength of
the tunnelling transition. To determine the spectrum of
Hamiltonian~(\ref{Ham1}) we take advantage of its periodicity. For that we
notice that the system repeats itself with respect to a unit cell that
comprises of two neighboring sites, e.g. along the ${\bf v}_3$ vector, as
seen in Fig.~\ref{fig:honeycomb}.
This periodic structure allows us to perform a Fourier transformation that
reduces the energy eigenvalue problem to the diagonalization of the
Hamiltonian in a single unit cell. Indeed, for $a({\bf k}) =
\sum_{\bf i} \, e^{i\kk \cdot \ii} a_{\ii}$ we obtain
\begin{equation}
H = -J\int\!\!\!\! \int d^2k \big(a_A^\dagger(\kk),
a_B^\dagger(\kk)\big)\left(
\begin{array}{cc}
0 & \sum_{i=1}^3 e^{i\kk \cdot {\bf v}_i} \\
\sum_{i=1}^3e^{-i\kk \cdot {\bf v}_i} & 0 \\
\end{array}
\right)
\left(
\begin{array}{c}
a_A(\kk) \\
a_B(\kk) \\
\end{array}
\right),
\label{Ham2}
\end{equation}
where $a_A(\kk)$ and $a_B(\kk)$ correspond to the Fourier transformed
operators that are positioned on the left and on the right of the unit cell,
respectively. Now, it is straightforward to find the eigenvalue of the
energy for electrons with certain momentum. This only requires diagonalizing
the two by two matrix inside the integral and yields
\begin{equation}
E({\kk}) =\pm J \sqrt{1 +4 \cos^2{\sqrt{3}k_y\over 2} + 4 \cos {3 k_x \over
2}\cos {\sqrt{3}k_y \over 2}}.
\label{energy}
\end{equation}
Common as they might look, these energy eigenvalues possess a unique
property. They become zero for isolated points of momentum. These are called
Fermi points and are positioned at ${\bf K}_\pm=2\pi/3(1,1/\sqrt{3})$. The
presence of Fermi points makes the low energy behavior of graphene so
special. At half fermionic filling and at zero external potential graphene
is at zero energy in the sense that the valence band is completely filled.
In this case the behavior of graphene is dominated by the behavior of the
Hamiltonian~(\ref{Ham2}) near the Fermi points.
\begin{figure}[h]
\begin{center}
\includegraphics[width=3.5cm]{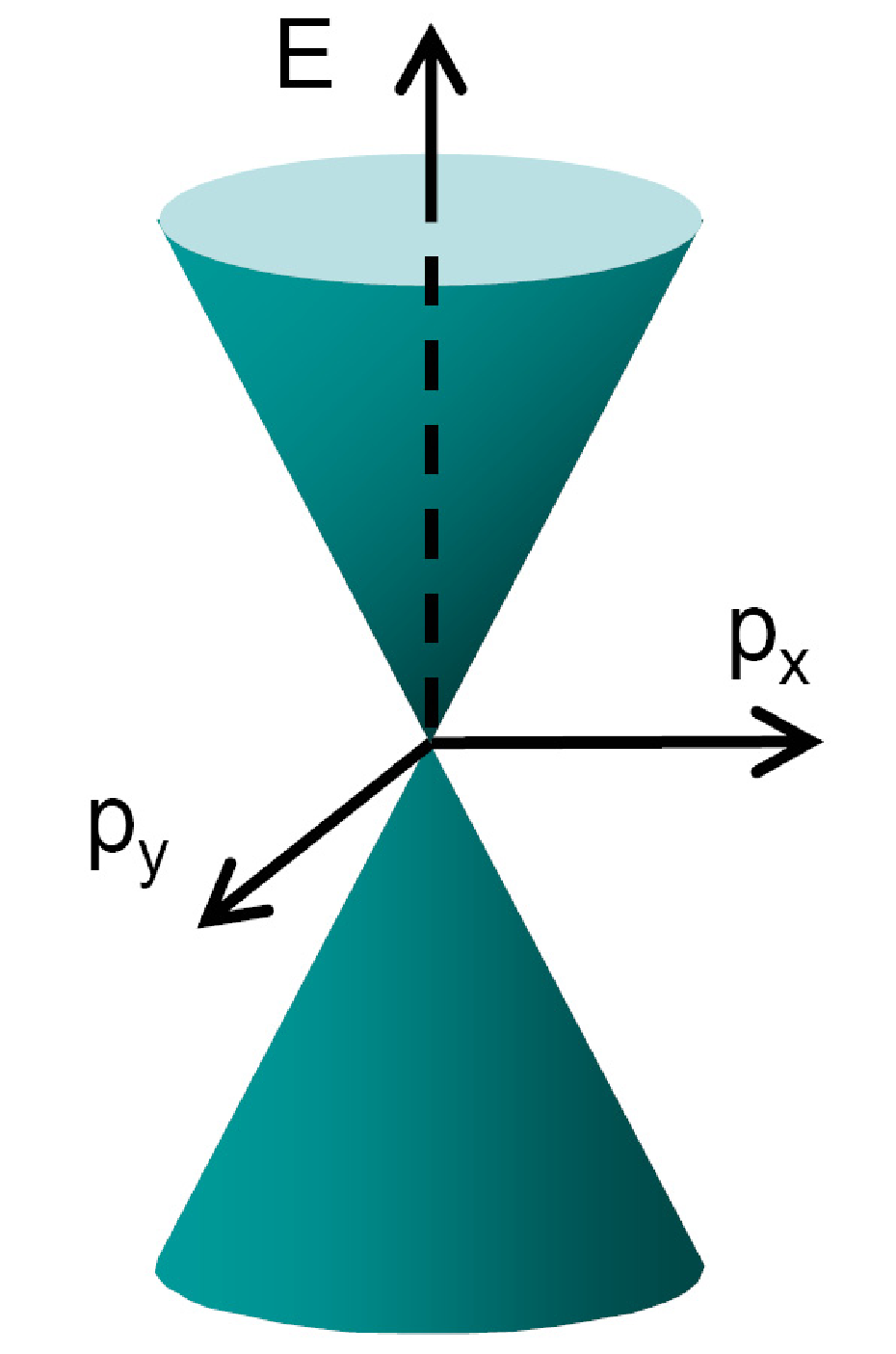}
\caption{The spectrum, $E$, of graphene near a single Fermi point as a
function of the small momenta, $p_x$ and $p_y$. The conical shape
corresponds to a linear relation between the energy and the momentum that
characterizes the Dirac equation. The positive spectrum corresponds to the
conducting band and the negative to the valence band.}
\label{fig:Dirac}
\end{center}
\end{figure}
To determine this behavior we substitute ${\bf k} ={\bf K}_\pm+{\bf p}$,
expand the Hamiltonian in powers of $|\p|$ and take into account only the
first order terms by assuming that $|\p|$ is small. This gives the equation
\begin{equation}
\left(
\begin{array}{cc}
H_+ & 0 \\
0 & H_- \\
\end{array}
\right)\Psi(\rr)=
\left(
\begin{array}{cc}
\SIGMA\cdot \p & 0 \\
0 & (\SIGMA\cdot \p)^* \\
\end{array}
\right)\Psi(\rr) = \GAMMA\cdot \p \Psi(\rr) =E \Psi(\rr),
\label{dirac}
\end{equation}
where $\p=-i\hbar \NABLA$, $\SIGMA=(\sigma^x,\sigma^y,\sigma^z)$ are the
Pauli matrices and $H_\pm$ is the Hamiltonian kernel that corresponds to
each Fermi point ${\bf K}_\pm$. Moreover, $\GAMMA$ are the Dirac matrices
given by
\begin{equation}
\GAMMA =
\left(
\begin{array}{cc}
\SIGMA & 0 \\
0 & -\SIGMA^* \\
\end{array}
\right)
\end{equation}
and for simplicity we have set $3J/2=1$. Eqn.~(\ref{dirac}) shows that, due
to the presence of isolated Fermi points, the low energy behavior of
graphene can be described by a relativistic Dirac operator,
$\slash\!\!\!\!D=\GAMMA\cdot\p$, with energy dispersion relations given in
Fig.~\ref{fig:Dirac}. The corresponding eigenvectors are four dimensional
spinors, $(\ket{{\bf K}_+A},\ket{{\bf K}_+B},\ket{{\bf K}_-A},\ket{{\bf
K}_-B})^T$, where the components are the wave functions of the unit cell
elements $A$ and $B$ for each of the Fermi points $\bf{K}_\pm$.

It is surprising that these effective Dirac fermions do not have any mass
term. This fact is responsible for the high mobility properties of graphene.
The presence of the Fermi points is responsible for the semi-metallic
behavior of graphene. That is, it behaves as metal due to the vanishing gap
between the valence and conducting bands, as seen in Fig.~\ref{fig:Dirac},
but with charge carriers in small numbers due to the vanishing density of
states at the Fermi level. Finally, there are emerging pseudo-spin degrees
of freedom, where the Pauli matrices are acting, as we initially started
with spinless electrons.

Of course, we have not discovered a fundamental relativistic field. We have
only shown that the collective behavior of graphene electrons can be
described by the Dirac equation. Indeed, the obtained effective velocity of
the electrons is three hundred times smaller than the speed of light, $c$.
The relativistic description of graphene's electrons gives rise to numerous
new phenomena that can be tested in the laboratory, like the Klein
paradox~\cite{Geim_klein}, the anomalous integer quantum Hall
effect~\cite{Hall} and so on.

\subsection{Curvature effect}

The next theoretical surprise is related to the way curved graphene is
described. As may be naturally expected the low energy description of a
curved graphene surface is given by a Dirac equation that is defined on the
corresponding curved surface. But this is not the full story. The curvature
induces an effective gauge field that needs to be introduced in the Dirac
equation. The latter is equivalent to an actual gauge field with magnetic
field flux going through the molecule surface. This is a remarkably simple
way to couple gauge fields to Dirac fermions.

\begin{figure}
\begin{center}
\includegraphics[width=3.5cm]{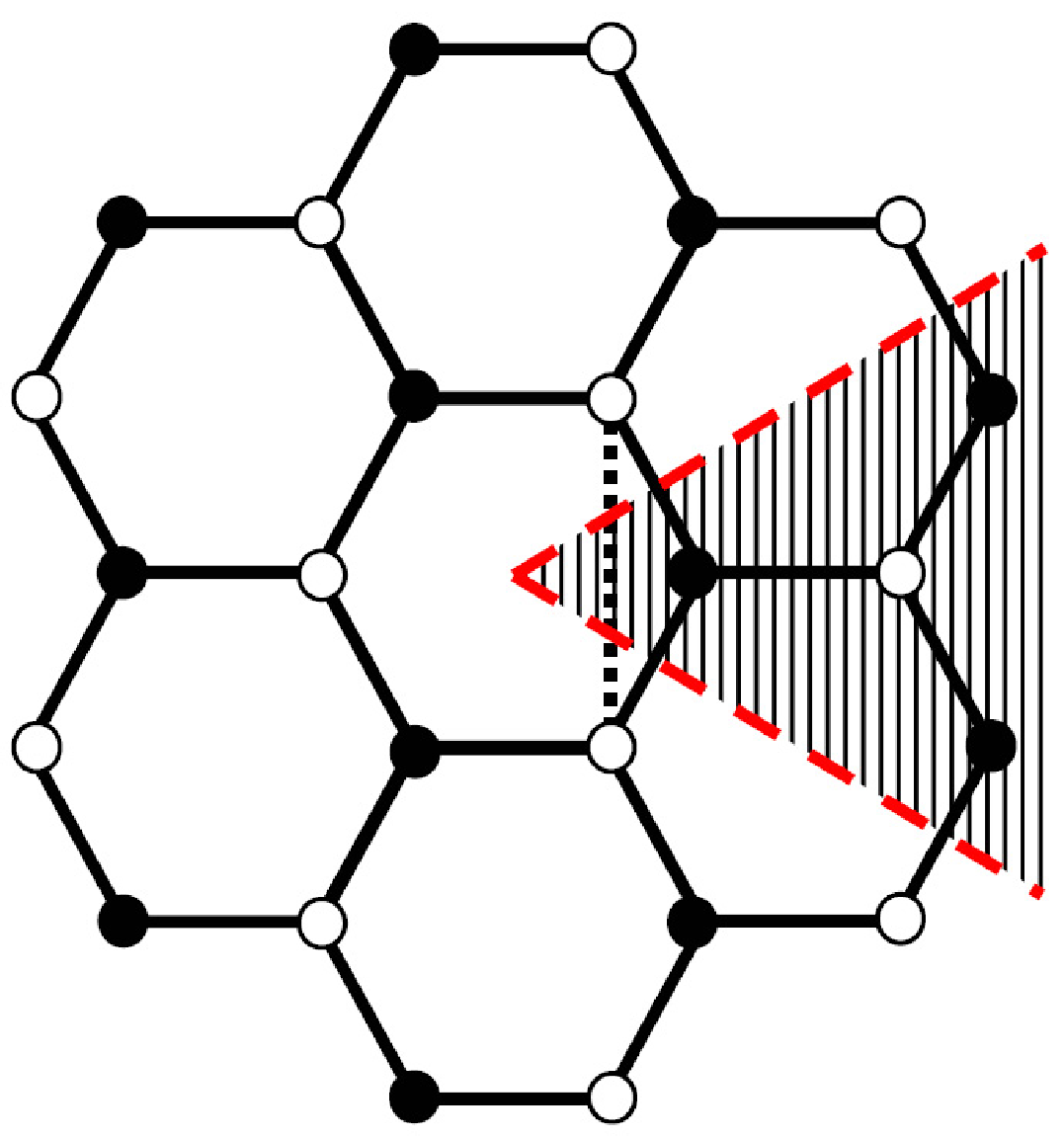}
\caption{The introduction of curvature on a flat sheet of
graphene by cutting a $\pi/3$ sector centered at the middle of a hexagon and
gluing the opposite ends. This, finally, creates a cone that has positive
curvature. The resulting lattice has a single pentagon at the apex of the
cone, while the rest of the plaquettes remain hexagonal.}
\label{fig:cone}
\end{center}
\end{figure}

To understand the mechanism that gives rise to the gauge fields let us
examine how one can insert curvature to graphene. The simplest way is by
cutting a $\pi/3$ sector from a graphene sheet and then gluing the opposite
ends of the lattice, as seen in Fig.~\ref{fig:cone}. This process adds a
single pentagon at the apex of the created cone. All the other plaquettes
remain hexagonal. This minimal geometrical distortion of the honeycomb
lattice adds positive curvature. The curvature can be measured by
circulating a tangential vector, ${\bf T}$, around the apex, i.e.
\begin{equation}
\oint {\bf T}\cdot d{\bf r} ={\pi \over 3}.
\end{equation}
But this is not yet the end of the story! The insertion of a single pentagon
forces us to connect two sites that are of the same type, e.g. $B$ for the
case in Fig.~\ref{fig:cone}. This is different compared to the flat
graphene, where every site $A$ has only $B$ neighbors or via versa. This
discontinuity has a dramatic effect on the corresponding spinor that
describes the deformed lattice. Indeed, when the spinor, $\Psi$, is
transported around the apex by an angle $2\pi$ it is forced at some point to
jump from a site $B$ to a site $B$ instead of a site $A$. This reminds us of
the effect that a magnetic field has on the wave function of an electron
moving on a closed path. A full circulation gives a phase factor to the
electron wave function that is proportional to the enclosed magnetic flux.
This effect is known as the Aharonov-Bohm effect~\cite{Aharonov}. Quantum
mechanically, the magnetic field introduces a discontinuity: the wave
function of the electron should simultaneously describe the processes of the
electron remaining static or spanning a closed trajectory. This is rectified
by introducing a vector potential term in the Hamiltonian that makes the
theory consistent. Following similar steps one needs to add a non-Abelian
vector potential term, ${\bf A}$, in the effective Hamiltonian description
of the curved graphene that compensates the jump in the components of the
spinor, $\Psi$. The circulation of ${\bf A}$ along a loop, $C$, around the
apex is given by
\begin{equation}
\oint_C {\bf A}\cdot d{\bf r} = {\pi \over 2} \tau^y,
\label{Acirc}
\end{equation}
where $\tau^y$ is the second Pauli matrix that mixes the ${\bf K}_+$ and
${\bf K}_-$ components of the spinor. This fascinating emergence of the
effective gauge field is simply caused by the geometric distortion of the
graphene lattice. The manifestation of the gauge potential is witnessed by
the properties of geometric variants of graphene, such as the spherical
fullerenes, and it is in the core of the topological effects described here.

The resulting Hamiltonian includes the curvature effect and a non-Abelian
gauge field that couples the Fermi points, ${\bf K}_+$ and ${\bf K}_-$. It
is possible to rotate this Hamiltonian so that the gauge field reduces to
two Abelian fields that act independently on each Fermi point. In this
rotated frame the effective Hamiltonian is given by
\begin{equation}
H=
\left(
\begin{array}{cc}
-ie^\mu_k\sigma^k (\nabla_\mu -ie A_\mu) & 0 \\
0 &  ie^\mu_k\sigma^k (\nabla_\mu +ie A_\mu)\\
\end{array}
\right).
\label{Ham_Dirac}
\end{equation}
The covariant derivative is given by $\NABLA=\PARTIAL -i\OMEGA$, where
$\OMEGA$ is the spin connection that describes the effect of curvature on
the vector character of the spinors. It has circulation
\begin{equation}
\oint_C \OMEGA\cdot d{\bf r} = -{\pi \over 6}\sigma^z,
\end{equation}
where $\sigma^z$ acts on the $A$ and $B$ components of the spinor and $C$ is
a loop around the apex of a cone. The Abelian gauge potential satisfies
\begin{equation}
\oint_C {\bf A}\cdot d{\bf r} = {\pi \over 2},
\end{equation}
with the corresponding magnetic field given by ${\bf B} =\NABLA\times{\bf
A}$. The zweibeins, $e_k^\mu$, are introduced to make the Hamiltonian
covariant in the induced curved space. If the curved space, with metric
$g^{\mu\nu}$, has coordinates $x^\mu$ and the local flat space, with metric
$\eta^{kl}$, has coordinates $\xi^k$ then the zweibeins are defined by
$e_k^\mu=\partial x^\mu/\partial\xi^k$ and they satisfy $g^{\mu\nu} =
e_k^\mu e_l^\nu\eta^{kl}$. Practically, the zweibeins transform the Pauli
matrices from the flat to the curved space in order to be contracted with
the corresponding covariant derivative. From the metric $g_{\mu\nu}$ one can
define the corresponding Christoffel symbols, $\Gamma_{\mu
\nu}^\sigma = {1
\over 2} g^{\sigma
\rho}(\partial_\mu g_{\nu
\rho} +\partial_\nu g_{\mu \rho}-\partial_\rho g_{\mu \nu})$,
while the scalar curvature is given by
\begin{equation}
R = g^{\mu\nu}R^\rho_{\,\,\mu\nu\rho}, \,\,\text{where}\,\,
R^\mu_{\,\,\,\nu\rho\sigma} =\partial_\sigma\Gamma_{\nu\rho}^\mu-
\partial_\rho\Gamma_{\nu\sigma}^\mu+
\Gamma_{\nu\rho}^\lambda\Gamma_{\lambda\sigma}^\mu-
\Gamma_{\nu\sigma}^\lambda\Gamma_{\lambda\rho}^\mu.
\end{equation}

The final Dirac equation, $H\Psi = E\Psi$, where $H$ is given in
(\ref{Ham_Dirac}), describes the behavior of a Dirac spinor, $\Psi$, coupled
to a gauge field, ${\bf A}$, on a surface with curvature $R$. This model
approximates well the low energy behavior of the carbon
molecules~\cite{DiVincenzo,Gonzalez2,Osipov,Kolesnikov,Lammert}, which
corresponds to states with large wavelength. The lattice spacing appears to
be small for these states and the continuous limit can be safely taken. To
be precise, the continuous limit corresponds to the limit of infinitely
large molecules, where states with arbitrarily large wavelength can be
supported. Nevertheless, we will see that the continuous approximation
describes graphene molecules well even for relatively small lattices.
Hamiltonian~(\ref{Ham_Dirac}) has all the necessary ingredients that are
responsible for the topological character of graphene molecules. It is now
of interest to see what specific form the Dirac equation takes for the
various geometric deformations of graphene.

\section{Geometric variants of graphene}

\begin{figure}
\begin{center}
\includegraphics[width=8cm]{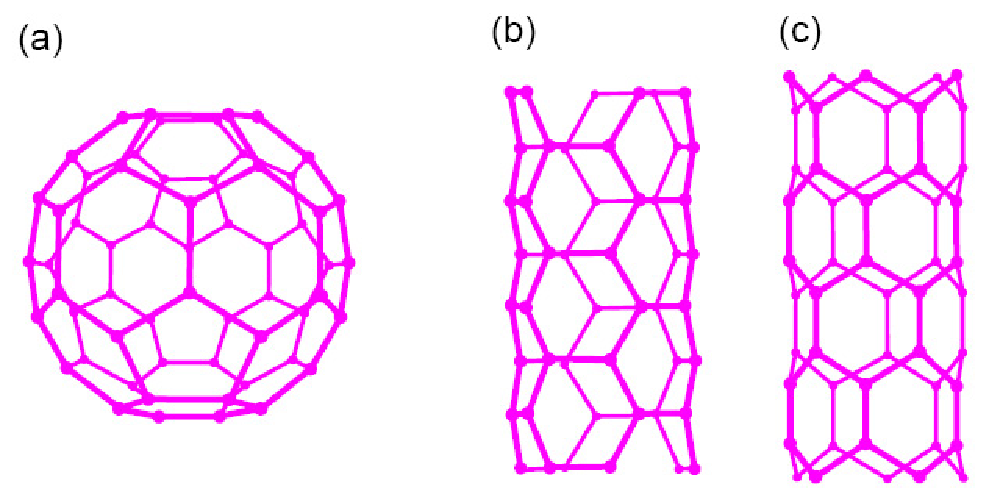}
\caption{(a) The C60 fullerene molecule, where there are 12 pentagons
necessary to produce the spherical configuration. Nanotube of armchair type
(b) and of zigzag type (c). It is not necessary to introduce pentagons for
producing nanotubes as they have zero curvature.}
\label{fig:fulenano}
\end{center}
\end{figure}

Among the most common geometrical variants of graphene are the
fullerene~\cite{Kroto} and the nanotubes~\cite{Radushkevich}. Fullerenes,
also known as buckyballs, are compact molecules with zero genus. As a
representative, the C60 has sixty carbon atoms positioned in a spherical
configuration, as seen in Fig.~\ref{fig:fulenano}(a). In order to create the
curvature of the sphere one has to introduce twelve pentagons on an
initially flat sheet of hexagons. This is the case for all buckyballs of
arbitrary size as we shall explicitly see in the next section. On the
contrary nanotubes are just folded graphene sheets in a cylindrical
configuration. Nanotubes have zero two dimensional curvature so there is no
need for the presence of any pentagons. Due to their lattice structure there
are many distinct ways in which a nanotube can be folded. Each one is
characterized by different boundary conditions. Examples are the armchair
(Fig.~\ref{fig:fulenano}(b)) and the zigzag configuration
(Fig.~\ref{fig:fulenano}(c)).

Of course, one can imagine more complex molecules with positive as well as
negative curvature. Negative curvature is present, for example, at saddle
points and can be introduced by heptagons. The latter introduce an effective
magnetic field flux that is exactly opposite to the one corresponding to
pentagons. By employing pentagons and heptagons, one can generate any
desirable graphene surface. These include compact surfaces of arbitrary
genus as well as open ones. In the following we consider only open surfaces
that can be produced by a normal section of compact ones. An example is the
nanotube that is produced from a torus by cutting it along its smallest
circle.

The Hamiltonian that describes the low energy spectrum of each of these
molecules is given by a Dirac equation similar to the one
in~(\ref{Ham_Dirac}). Now the metric, $g_{\mu\nu}$, describes the curvature
of the surface of each molecule. The gauge field, ${\bf A}$, is responsible
for the local magnetic flux through every polygonal deformation. Thus, by
keeping track of the geometry of the surface of the molecule as well as its
polygonal deformations one is able to construct the corresponding Dirac
operator that describes its low energy limit. Tedious as it may seem there
is a simple way to determine the total flux going through the molecule from
its general topological characteristics. This is achieved with the aid of
the Euler theorem, which is presented in the following section.

\section{Euler theorem}

As we have seen, to establish the low energy description of graphene
molecules we need to determine the total number of plaquette deformations.
If we know the total number of pentagons and heptagons present in a molecule
of certain genus, then we can determine the total flux that goes through the
molecule. There is a simple way to relate the number of these deformations
to the topology of the corresponding molecule. Indeed, the Euler
theorem~\cite{Lakatos} gives a relation between the structural details of a
polyhedral lattice and its total topological properties. There are numerous
proofs of this theorem, the most pedagogical ones are based on a reduction
of the polyhedral lattice to simpler ones without changing its topological
properties.

\begin{figure}[h]
\begin{center}
\includegraphics[width=9cm]{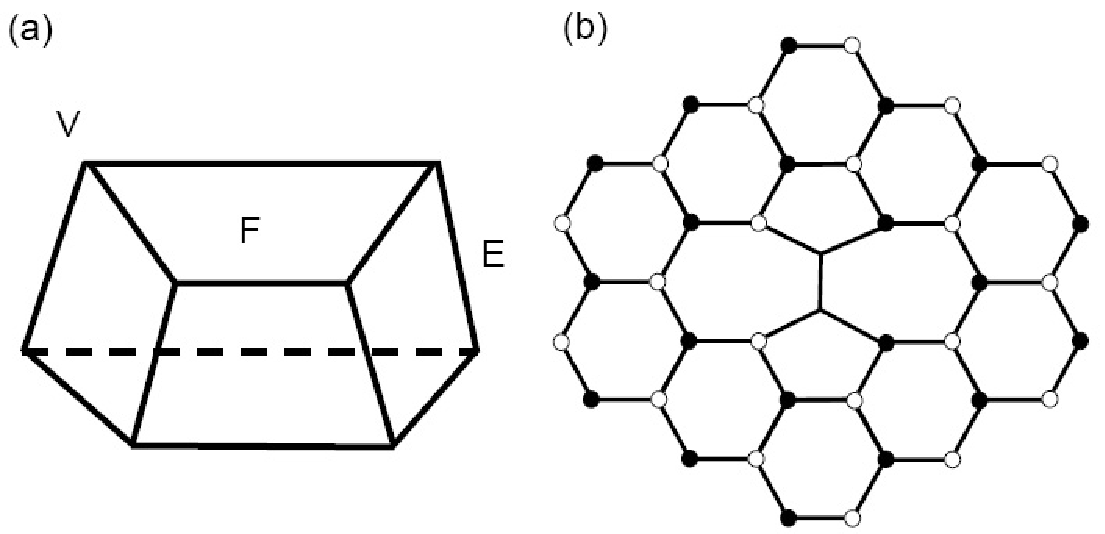}
\caption{(a) A simple lattice corresponding to genus zero surface,
where the vertices, $V$, edges, $E$, and faces, $F$, are depicted. (b) Two
pentagonal and two heptagonal deformation can be introduced on graphene
lattice without changing its flat geometry.}
\label{fig:Euler}
\end{center}
\end{figure}

Consider a lattice that lies on a surface that we initially take to be
compact of a certain genus, $g$. An example of such a lattice with genus
$g=0$ is seen in Fig.~\ref{fig:Euler}(a). Let $V$, $E$ and $F$ be the number
of vertices, edges and faces of the lattice, respectively. The Euler
characteristic, $\chi$, is given by
\begin{equation}
\chi = V-E+F = 2(1-g),
\label{Euler1}
\end{equation}
where the second equation consists the Euler theorem. We now apply this
theorem to the case of graphene molecules. Each graphene vertex has exactly
three links. Let us assume that only pentagonal or heptagonal deformations
are present. Denote by $n_5$, $n_6$ and $n_7$ the total number of pentagons,
hexagons and heptagons in the molecule. Then the total number of faces is
given by the sum of the different polygons, $F= n_5+n_6+n_7$. The total
number of vertices is given by $V=(5n_5+6n_6+7n_7)/3$ as each $k$-gon has
$k$ vertices and each vertex is shared by three polygons. Similarly, the
total number of edges is given by $E =(5n_5+6n_6+7n_7)/2$ as each edge is
shared by two polygons. Substituting this into Eqn.~(\ref{Euler1}) we obtain
\begin{equation}
n_5-n_7= 12(1-g).
\label{Euler_gr}
\end{equation}
There are many characteristics that spring out from this result. If equal
numbers of pentagons and heptagons are inserted then they do not change the
topology of the surface as they cancel out. Indeed, on a flat graphene sheet
one can add two pentagons and two heptagons without changing the curvature
of the molecule away from these deformations, as seen in
Fig.~\ref{fig:Euler}(b). This is consistent with the effective gauge field
description, where pentagons and heptagons have opposite flux contributions.
On the other hand, non-trivial topologies necessarily introduce an imbalance
between pentagons and heptagons. The genus zero configurations need an
excess of pentagons, while high genus surface need an excess of heptagons.
Genus one surfaces do not need any pentagons or heptagons at all as they are
equivalent to a flat sheet.

As particular examples we see that Eqn.~(\ref{Euler_gr}) reproduces the
known case of a sphere with $g=0$ that gives $\chi=2$. This corresponds to
the fullerenes that have $n_5=12$ and $n_7=0$. For the case of the torus we
have $g=1$ giving $\chi=0$. Thus, no pentagons or heptagons are required. If
one wants to consider the genus $g=2$ surfaces then $\chi=-2$ for which
$n_5=0$ and $n_7=12$. Fullerenes are compact surfaces, while the rest of the
geometrical variants of graphene are usually defined on open surfaces. To
apply our study to them we need to insert open boundary conditions to the
Euler characteristic. This is simply achieved by considering the
corresponding compact surface and introducing normal cuts to it. As an
example, a torus with $g=1$ can be cut along its small circle thus creating
a cylinder with two open faces. The type of polygons between these two
shapes is the same. So the generalization of the Euler theorem gives
\begin{equation}
n_5-n_7= 12(1-g)-6N,
\label{euler1}
\end{equation}
where $N$ is the number of open faces. The surgical procedure presented
above, employed to produce open faces gives two open faces when the genus is
reduced by one. In the case of nanotubes the surface has $g=0$ and $N=2$
that corresponds to $n_5=n_7=0$. The case with $g=0$ and $N=4$ corresponds
to two crossing nanotubes. In general, arbitrary configurations of genus $g$
and of open faces $N$ can be considered. Then the imbalance between the
polygonal deformations is determined by relation (\ref{euler1}). It remains
to see how the total flux associated with the total number of pentagons and
heptagons can determine the low energy spectrum of the molecule. In the next
section we see how the index theorem can provide such an elegant relation.

\section{The Atiyah-Singer index theorem}

The index theorem gives an insight into the structure of the spectrum of
certain operators, like the Dirac operator, without having to diagonalize
them~\cite{Atiyah,Eguchi,Stone}. This information can be derived from
general properties of the operators and the geometry of the space in which
they are defined. In particular, the theorem is concerned with eigenstates
that have zero eigenvalue, which in our case are the zero modes. Abstract as
it might first sound it is not hard to follow a heuristic proof of the index
theorem. Here we employ the approach based on the heat kernel expansion that
provides the main elements, applicability and limitations of the theorem.
The index theorem allows us to obtain a relation between the zero modes of
the Dirac operator that corresponds to graphene and the total flux that goes
through its surface. As the latter is related to the genus of the surface
through the Euler characteristic we finally obtain, as promised, a relation
between the zero modes and the topology of the surface.

Our starting point is a Dirac operator of the form
\begin{equation}
\slash\!\!\!\!D=\left(
\begin{array}{cc}
0 & P^\dagger \\
P & 0 \\
\end{array}
\right),
\end{equation}
where $P$ is an operator that maps a space $V_+$ onto a space $V_-$, while
$P^\dagger$ maps $V_-$ onto $V_+$. If $P$ is an $n\times m$ matrix, then
$P^\dagger$ is a $m\times n$ matrix and $V_+$, $V_-$ is the space of $n$,
$m$ dimensional vectors, respectively. For the Hamiltonian~(\ref{Ham_Dirac})
obtained in the previous section, $\slash\!\!\!\!D$ is one of the two
sub-blocks of the diagonal differing by the sign in front of the gauge
field, ${\bf A}$, $P$ is a differential operator and $V_\pm$ is the space of
complex functions.

As we are interested in the zero modes of $\slash\!\!\!\!D$, i.e. the
solutions of the equation $\slash\!\!\!\!D\Psi = 0$, we can define the
number of distinctive eigenstates of $P$ with zero eigenvalue as $\nu_+$ and
the ones of $P^\dagger$ as $\nu_-$. To facilitate the bookkeeping we
introduce the chirality operator $\gamma_5$ as
\begin{equation}
\gamma_5 =
\left(
\begin{array}{cc}
1 & 0 \\
0 & -1 \\
\end{array}
\right).
\end{equation}
Its eigenstates have eigenvalue $\pm1$ if they belong to $V_\pm$. To
simplify our calculations we consider the operator $\slash\!\!\!\!D^2$ that
has the same number of zero eigenstates as $\slash\!\!\!\!D$, but has the
additional advantage of being diagonal
\begin{equation}
\slash\!\!\!\!D^2=\left(
\begin{array}{cc}
P^\dagger P & 0 \\
0 & P P^\dagger \\
\end{array}
\right).
\end{equation}
It is possible to show that the operators $P^\dagger P$ and $P P^\dagger$
have the same non-zero eigenvalues. Indeed, if there is a state $u$ such
that $PP^\dagger u = \lambda u$, for eigenvalue $\lambda\neq 0$, then
\begin{equation}
PP^\dagger u = \lambda u \Rightarrow P^\dagger P(P^\dagger u) = \lambda
(P^\dagger u),
\label{symmetry}
\end{equation}
which means that also the operator $P^\dagger P$ has the same eigenvalue,
$\lambda$, corresponding to the eigenstate $P^\dagger u$. However, this is
not necessarily the case for $\lambda=0$ as $P^\dagger u$ could be zero by
itself.

Let us now evaluate the following quantity
\begin{equation}
\text{Tr} (\gamma_5e^{-t\slash\!\!\!\!D^2}) =
\text{Tr}(e^{-t P^\dagger P})- \text{Tr} (e^{-tPP^\dagger}) =
\sum_{\lambda_+} e^{-t\lambda_+} -\sum_{\lambda_-} e^{-t\lambda_-},
\label{trace}
\end{equation}
where $\lambda_+$ and $\lambda_-$ are the eigenvalues of the operators
$P^\dagger P$ and $PP^\dagger$, respectively, and $t$ is just an arbitrary
parameter. The first step in~(\ref{trace}) is due to $\gamma_5$ acting on
the exponential: it contributes a $+1$ to the eigenvectors of $P^\dagger P$,
as they belong to $V_+$, and a $-1$ to the eigenvectors of $PP^\dagger$, as
they belong to $V_-$. In the last step we evaluated the trace as a sum over
all the eigenvalues of the corresponding operators. As we have shown, every
non-zero eigenvalue of $P^\dagger P$ corresponds to an eigenvalue of
$PP^\dagger$. Thus, all terms with non-zero eigenvalues cancel out in pairs
and only the zero eigenvalues of each operator remain, finally giving
\begin{equation}
\text{Tr} (\gamma_5e^{-t\slash\!\!\!\!D^2}) = \nu_+-\nu_-.
\label{index1}
\end{equation}
The difference between the number of zero modes is in general undetermined.
We call it the index of the $\slash\!\!\!\!D$ operator, i.e.
$\text{index}(\slash\!\!\!\!D)
\equiv
\nu_+-\nu_-$. The above derivation provides us with an unexpected clue: due to the
cancellation of the non-zero eigenvalue terms in~(\ref{trace}) the
expression $\text{Tr} (\gamma_5e^{-t\slash\!\!\!\!D^2})$ is actually $t$
independent!

What we just presented is merely a definition. Here, we are actually
interested in finding out what is the value of
$\text{index}(\slash\!\!\!\!D)$. For that we shall employ an alternative way
of calculating $\text{Tr} (\gamma_5e^{-t\slash\!\!\!\!D^2})$ called the heat
kernel expansion method~\cite{Vassilevich}. It states that for a two
dimensional compact surface and for general $\hat f$ and $\hat D$ operators
one has the expansion
\begin{equation}
\text{Tr}(\hat f e^{-t \hat D}) = {1 \over 4\pi t} \sum_{k\geq
0}t^{k/2}a_k(\hat f, \hat D),
\label{heat}
\end{equation}
where $\text{Tr}$ denotes the trace of matrices as well as the integration
with respect to spacial coordinates and $a_k$ are expansion coefficients.
For $\hat f =\gamma_5$ and $\hat D=\slash\!\!\!\!D^2$ we should
obtain~(\ref{index1}), that is an expression which is $t$ independent. To
achieve this, the coefficients $a_k$ should be zero for all $k$ except for
$k=2$ for which all the $t$ contributions in (\ref{heat}) cancel out. The
value of $a_2$ can be found from the first order term in the $t$ expansion
of the exponential. For this calculation we first evaluate that $
\slash\!\!\!\!D^2 = -g^{\mu\nu}\nabla_\mu\nabla_\nu+{i\over
4}[\gamma^\mu,\gamma^\nu]F_{\mu\nu}- {1 \over 4} R$, where $F_{\mu \nu} =
\partial_\mu A_\nu-\partial_\nu A_\mu$ is the field strength and
$\nabla_\mu$ is a covariant derivative with respect to gauge and
reparametrization transformations. The magnetic field is given in terms of
the field strength by $B_k ={1\over 2} \epsilon^{k\mu\nu} F_{\mu\nu}$. One
can now calculate that the non-zero expansion coefficient, $a_2$, is given
by
\begin{equation}
a_2 = \text{Tr}\Big\{\gamma_5\big({i \over 4} [\gamma^\mu,\gamma^\nu]F_{\mu
\nu}-{1 \over 4} R\big)\Big\} =2\int \!\!\!\! \int {\bf B}\cdot d{\bf S}.
\end{equation}
The integration in this equation runs over the whole surface. Combining the
results from the two independent calculations of $\text{Tr}
(\gamma_5e^{-t\slash\!\!\!\!D^2})$ we have
\begin{equation}
\text{index}(\slash\!\!\!\!D) =\nu_+-\nu_-=
{1 \over 2 \pi}\int \!\!\!\! \int {\bf B}\cdot d{\bf S}.
\label{index2}
\end{equation}
This is the final formula that demonstrates the index theorem. It relates
the total flux that goes out of the surface to the number of zero modes of
the $\slash\!\!\!\!D$ operator. The curvature, $R$, does not appear in the
index expression as $\gamma_5$ is a traceless operator. The absence of
curvature is a characteristic of two dimensional surfaces. Finally, it is
worth noting that Eqn.~(\ref{index2}) makes sense only if the right hand
side is an integer. This is indeed the case when we consider compact
surfaces. Then $\int \!\! \int {\bf B}\cdot d{\bf S}$ gives the total
magnetic monopole charge inside the surface which takes discrete values due
to Dirac's quantization condition of monopoles~\cite{Coleman}. In the case
of non-compact surfaces the theorem is applicable only when there is no flux
going through the open faces.

\section{Spectra of graphene molecules}

In order to apply the index theorem to graphene and its geometrical variants
we have to determine the contribution of the effective magnetic field to
(\ref{index2}). As we have seen the effective magnetic field that appears in
a certain graphene configuration, with minimally distorted lattices, is
directly related to the genus of the graphene surface. In particular, we
know that each pentagonal distortion in the honeycomb lattice gives rise to
a specific circulation of the vector potential, namely $\oint_{C_p} {\bf
A}\cdot d{\bf r} = {\pi / 2} $, and similarly for heptagonal distortions. By
employing Stokes's theorem we can relate the circulation of the gauge
potential around a plaquette to the flux of the corresponding magnetic
field. Indeed, we have
\begin{equation}
\oint_{C_p} {\bf A}\cdot d{\bf r}= \int\!\!\!\! \int_{S_p} {\bf B}\cdot
d{\bf S},
\label{Stokes}
\end{equation}
where $C_p$ is the looping trajectory for which the circulation of ${\bf A}$
is evaluated and $S_p$ is the corresponding area. Thus, the total flux that
goes trough the surface of the molecule can be evaluated if we know the
total number of pentagons and heptagons. As heptagons contribute the
opposite flux than pentagons we are only interested in the difference of
their numbers. This is exactly determined by the genus of the surface
through the Euler characteristic, i.e. $6\chi = n_5-n_7 = 12(1-g)-6N$. The
total flux of the gauge field that goes through the surface of the molecule
is, hence, given by
\begin{equation}
{1 \over 2\pi}\int\!\!\!\! \int_{S} {\bf B}\cdot d{\bf S} = {1 \over
2\pi}\sum_{n_5-n_7}
\oint_{C_p} {\bf A}\cdot d{\bf r} =
{1 \over 2\pi}{\pi \over 2}(n_5-n_7) = 3(1-g)-{3 \over 2} N.
\end{equation}
The total number of zero modes is the sum of the contributions from each
sub-operator of $H$ in (\ref{Ham_Dirac}). Consequently, we obtain the index
of the Hamiltonian, $H$, that describes the molecule by adding these
contributions. This finally gives
\begin{equation}
\text{index}(H) = \nu_+-\nu_- = 6(1-g)-3N.
\label{final}
\end{equation}
This is our final result that relates the number of zero modes present in a
certain graphene molecule to the topological characteristics of its surface.
Eqn.~(\ref{final}) gives us the minimum number of possible zero modes. The
exact number is obtained from the index if either $\nu_+=0$ or $\nu_-=0$.
This actually holds for most of the cases when the index is non-zero.

We can now compare our results with the known cases of graphene molecules.
For example, a fullerene has $g=0$ and $N=0$. Thus, it is expected to have
six zero modes. This is indeed the case as the fullerenes have two triplet
modes with energies very close to zero. For the case of a nanotube we have
$g=0$ and $N=2$. This gives rise to index zero, which is in agreement with
previous theoretical and experimental results~\cite{Reich,Saito}. Nanotubes
have either zero modes that satisfy $\nu_+=\nu_-$ as happens for the
armchair nanotubes (see Fig.~\ref{fig:fulenano}(b)) or no zero modes at all,
as it is the case for the zig-zag nanotubes (see
Fig.~\ref{fig:fulenano}(c)). The balance between the chiral zero modes is a
consequence of the symmetry between the two directions along the nanotube
that forces the two chiralities to be symmetric. Finally, the number of zero
modes of more complicated molecules can be obtained from Eqn.~(\ref{final}).

\begin{figure}[h]
\begin{center}
\includegraphics[width=5cm]{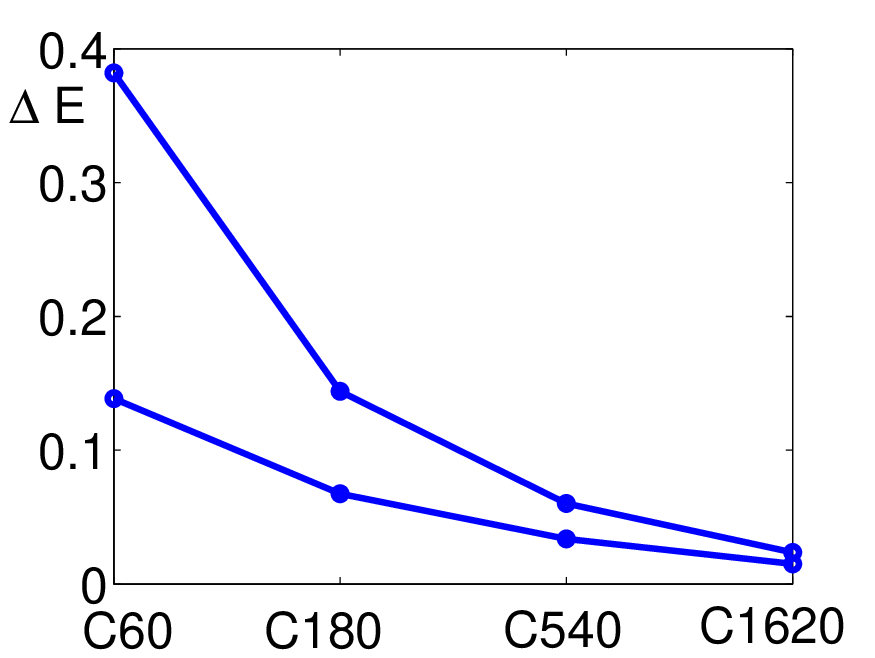}
\caption{The low lying energy levels for the fullerenes
C60, C180, C540 and C1620 obtained by direct diagonalization of their
lattice Hamiltonians. Two triplets are obtained that correspond to the six
zero energy modes in the continuum limit. As the size of the fullerene
increases the energies of the low lying modes tend rapidly close to zero.
The rest of the energy levels are separated with an energy gap of order one
above and below the depicted energies. The energies are in units of the
tunnelling coupling, $J$.}
\label{fig:lowmodes}
\end{center}
\end{figure}

When comparing this to the spectra of real molecules one has to realize that
small deviations are possible. Indeed, we have employed the continuous
approximation in our derivation, while actual graphene molecules are defined
on a lattice. Strictly speaking their lattice nature precludes them from
having exact zero modes. Nevertheless, we expect to obtain low lying states
that tend to zero modes when the molecule is taken to be larger and larger,
thereby approximating the continuous limit better and better. This is
clearly shown in Fig.~\ref{fig:lowmodes}, where the low lying spectra of the
fullerenes C60, C180, C540 and C1620 are plotted. For each molecule two
triplets are obtained corresponding to the six zero modes. This is as
expected in the continuous limit with energies that approach rapidly zero as
the size of fullerene increases. The rest of the spectrum is well separated
from these low lying states.

This is a surprising relation between the topology and the spectrum of
graphene molecules. In particular, it suggests that the zero modes need
certain conditions in order to exist. An important ingredient is the
presence of magnetic flux that is effectively inserted in graphene molecules
by geometrical deformations. By inducing curvature with minimal distortions
of the underlying lattice in the form of pentagons and heptagons we were
able to relate the number of these deformations to the general topological
characteristics of the lattice surface. This resulted in Eqn.~(\ref{final})
that relates the zero modes of a general graphene molecule with the genus
and the number of open faces of its surface.

\section{The next step: Vortices}

The previously described topological effects are not the only ones that can
be encountered in graphene. Recently, an inspiring relation between
vortex-like topological defects and the number of zero modes was
demonstrated~\cite{Chamon,Jackiw,Chamon1,Pachos3,Ghaemi,Franz}. This
construction is not very different from the one we studied in the previous
section. The crucial new element is the presence of a scalar field in the
Dirac operator on which vortices can be encoded. This scalar field is
generated in the effective picture by appropriate distortions of the
tunnelling couplings of the graphene molecules. The main characteristic of
the vortices is the presence of magnetic flux attached to them. The magnetic
field associated with the vortices can be, in principle, introduced by
external sources. Here we focus on the case where the field is generated by
the geometrical deformations that we already introduced in the previous
sections. This is a surprisingly simple mechanism compared to introducing
external fields.

\begin{figure}[h]
\begin{center}
\includegraphics[width=3.5cm]{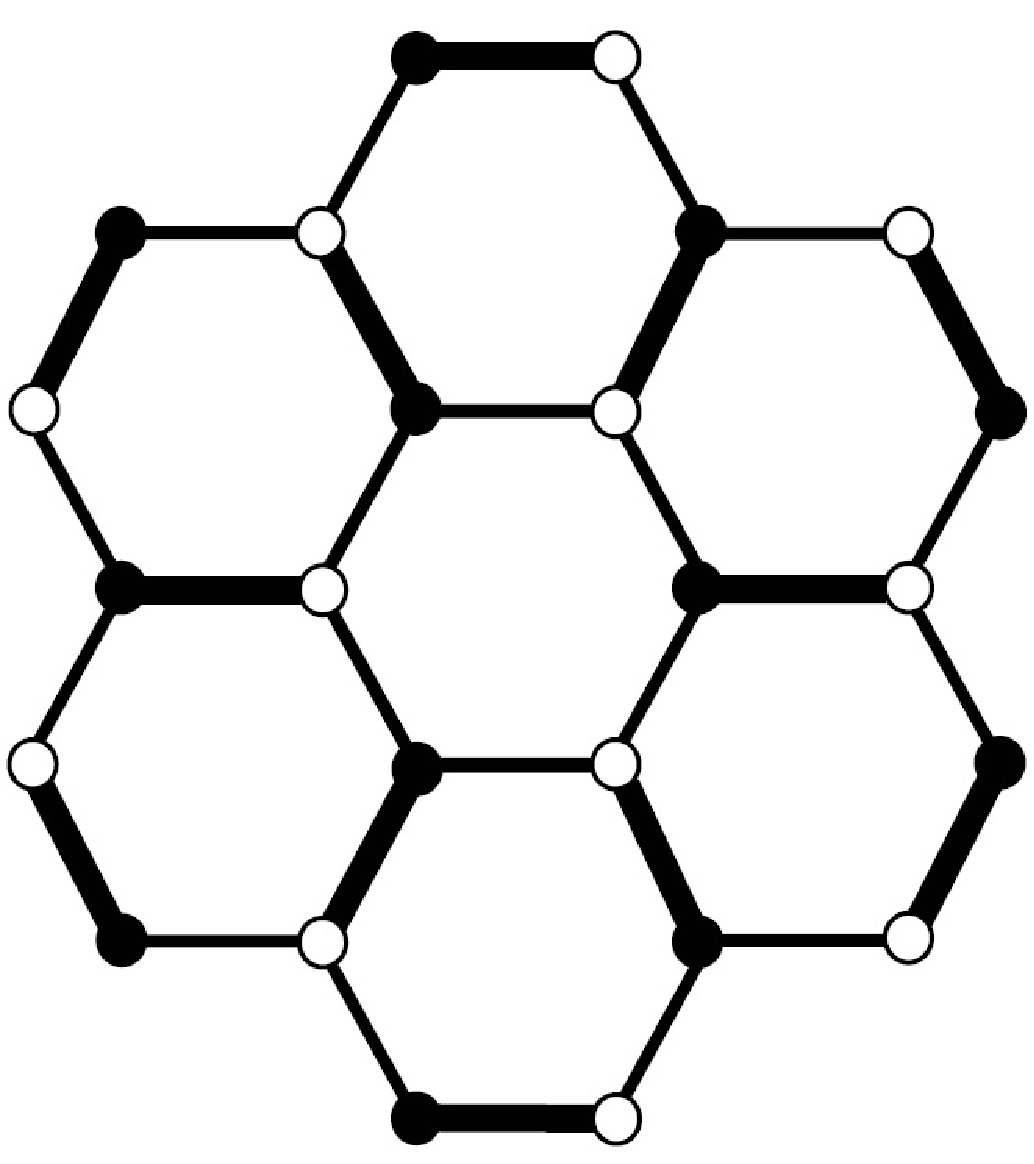}
\caption{The Kekul\'e distortion introduced on the honeycomb lattice.
A third of the hexagonal plaquettes has all the tunnelling couplings the
same, while the rest two thirds of the hexagons have alternating strong and
weak couplings.}
\label{fig:Kekule}
\end{center}
\end{figure}

Let us first see how the scalar field can appear in the effective low energy
behavior of graphene. Consider a particular distortion of the tunnelling
couplings of graphene, namely the Kekul\'e distortion, that is given by
\begin{equation}
\delta J_{{\bf r},k} = {1 \over 3} \Phi({\bf r})
e^{i{\bf K}_+\cdot {\bf v}_k} e^{i({\bf K}_+-{\bf K}_-)\cdot {\bf r}}
+\text{c.c.},
\end{equation}
where $\Phi$ is a slowly varying function, $K_\pm$ are the Fermi points,
${\bf r}$ gives the position of the lattice sites and ${\bf v}_k$, for
$k=1,2,3$, indicate the positions of the three neighboring sites, as
depicted in Fig.~\ref{fig:honeycomb}. It creates a pattern of weak, $J$, and
strong, $J'$, couplings as represented in Fig.~\ref{fig:Kekule}. The effect
of the distortion in the Dirac equation is to introduce a complex field,
$\Phi$, in the following way
\begin{equation}
H=
\left(
\begin{array}{cc}
-ie^\mu_k\sigma^k (\nabla_\mu -ie A_\mu) & \Phi \\
\Phi^* &  ie^\mu_k\sigma^k (\nabla_\mu +ie A_\mu)\\
\end{array}
\right),
\label{Ham_Scalar}
\end{equation}
where $\Phi = |\Phi|e^{i2e\int {\bf A}\cdot d{\bf r}}$. As we see, the
scalar field is directly coupled to the gauge field. The vorticity of the
scalar field is defined as
\begin{equation}
\text{Vort}(\Phi)\equiv
\frac 1{2\pi} \oint_{\partial \Omega} d{\bf r}\cdot \PARTIAL{\rm Arg\,}
\Phi ={1 \over 4\pi i}\oint_{\partial \Omega} d{\bf r} \cdot
\frac{\Phi^* {\PARTIAL} \Phi- \Phi{\PARTIAL} \Phi^*}{|\Phi|^2}.
\label{vorticity}
\end{equation}
It was argued by Jackiw and Rossi~\cite{Jackiw_81} and then proven by E.
Weinberg~\cite{Weinberg,Callias} that the index, which gives the zero modes
of the Hamiltonian~(\ref{Ham_Scalar}), is equal to the vorticity of the
scalar field, i.e. $\text{index}(H)=\text{Vort}(\Phi)$. An extension of this
theorem to compact surfaces was given in~\cite{Pachos3}. A direct
calculation of the vorticity of the scalar field gives us
\begin{equation}
\text{index}(H)= {1\over\pi}\int\!\!\!\! \int_{S} {\bf B}\cdot d{\bf S}.
\label{index3}
\end{equation}
Note, that this index is with respect to four dimensional Hamiltonians in
contrast to the pair of two Dirac equations we encountered in the previous.
The index is now twice the previous one reflecting the fact that the scalar
field has twice the charge of the fermions, $2e$. The contour $S$ is taken
to enclose all vortices imprinted on the scalar field.

Chemists have long surmised that in fullerenes not all the nearest-neighbor
hopping couplings are equal. In numerical calculations it is found that the
molecules can lower their electronic energy by undergoing a Kekul\'e
distortion. In particular, in the case of ``leapfrog" fullerene
molecules~\cite{leapfrog1} $C_{60+6k}$ with $k=0,1,...$ it is possible to
assign a Kekul\'e distortion in a consistent way. Thus, in addition to the
gauge field that corresponds to the pentagons one obtains a scalar field
from the possible coupling distortion.

From the previous study we know that the total effective magnetic flux of
fullerenes is $3\cdot 2\pi$. This property remains unchanged when a Kekul\'e
distortion is introduced. The equality between vorticity (\ref{vorticity})
and the corresponding index (\ref{index3}) assign to the scalar field a
total winding number of six. This is composed out of twelve half vortices
attached to the twelve pentagons of the carbon lattice. The vorticity $\pi$
of each vortex, instead of the flux $\pi/2$ of each pentagon, is due to the
charge $2e$ that corresponds to the scalar field.

The index theorem applied to fullerenes with coupling distortion seems
different than the one we employed for the undistorted molecules. However,
both give the same total number of zero modes. This underlines the fact
that introducing a scalar field in a Dirac operator does not change its
total index. The only extra complication that arises is the angular
discontinuity of the scalar field around the vortices due to their $\pi$
vorticity. This forces us to introduce branch cuts, i.e. lines between pairs
of vortices, where the scalar field takes zero value. Moreover, regularity
of the vortices demands that the scalar field is smooth around the vortices.

\begin{center}
\begin{figure}[ht]
\resizebox{13cm}{!}
{\hspace{12cm}\includegraphics{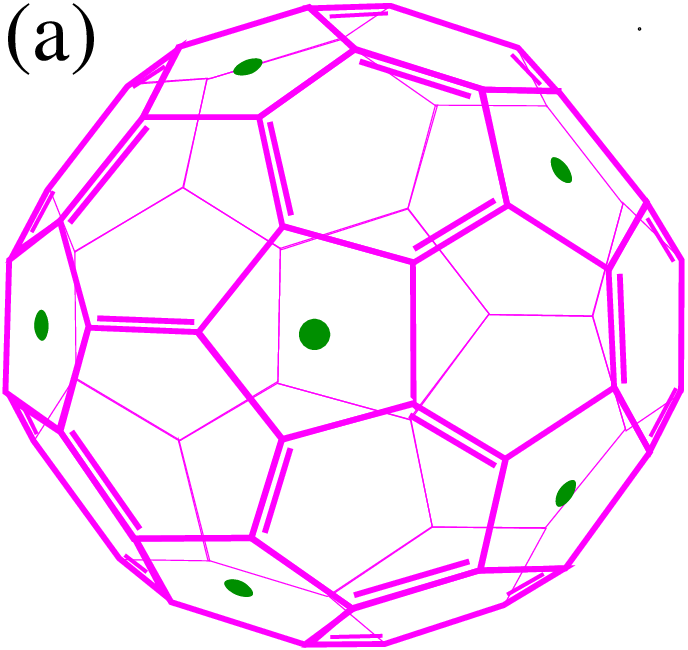}
\hspace{0.5cm}
\includegraphics{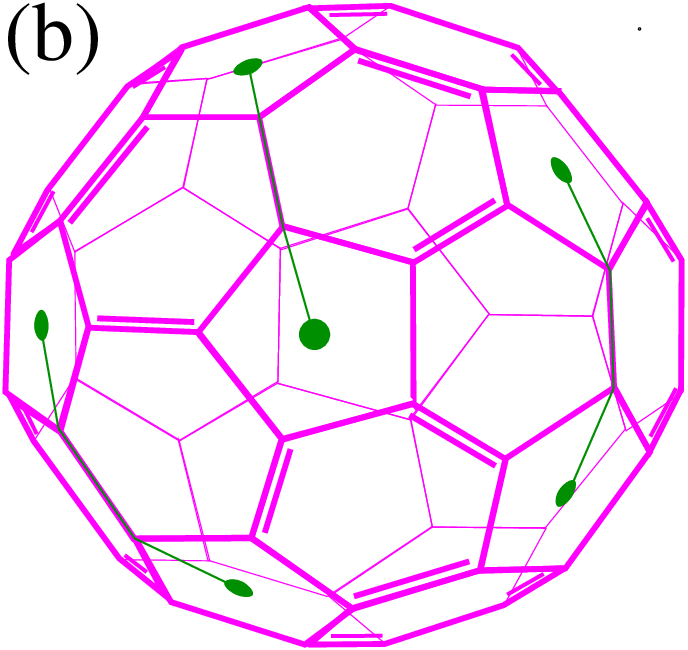}
\hspace{0.5cm}
\includegraphics{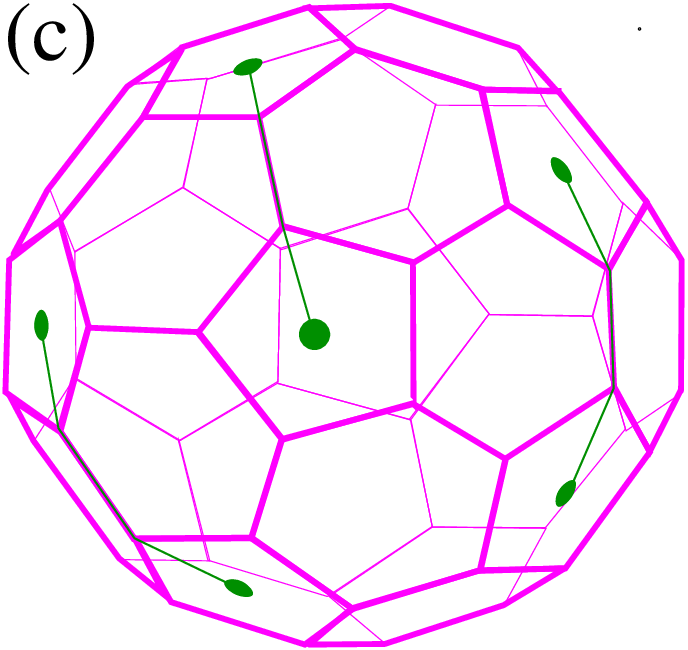}
 }
\caption{\label{C60Kekule} The coupling configuration of the C60
molecule, where vortices reside on the pentagons. (a) The Kekul\'e
distortion. (b) Cuts between vortices are introduced by replacing a double
bond with single ones. (c) An enlargement of the vortices is introduced by
removing all double bonds connected to the pentagons. For C60 this removes
all double bonds.}
\end{figure}
\end{center}

\begin{center}
\begin{figure}[h]
\resizebox{14cm}{!}
{\hspace{8cm}
\includegraphics{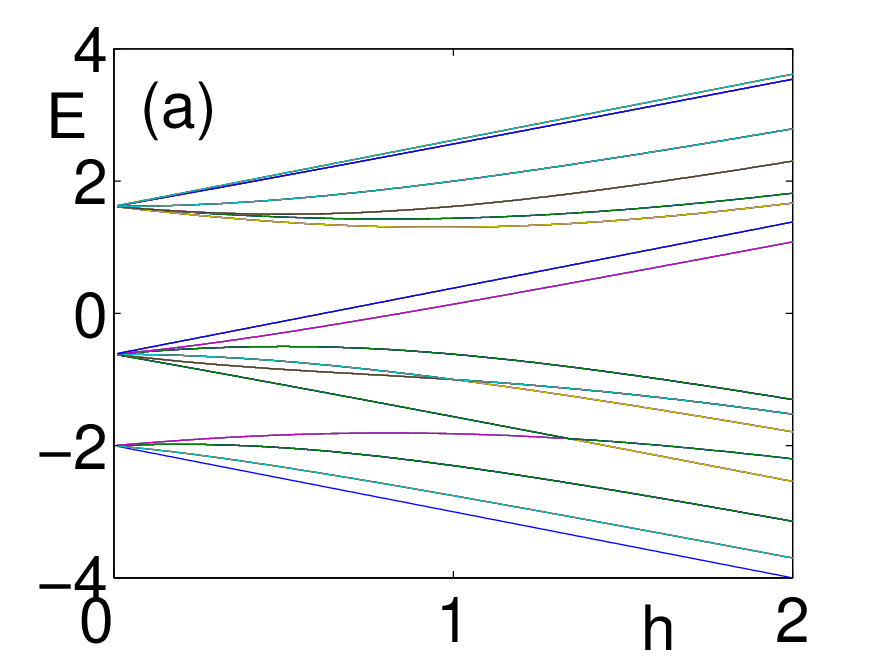}
\includegraphics{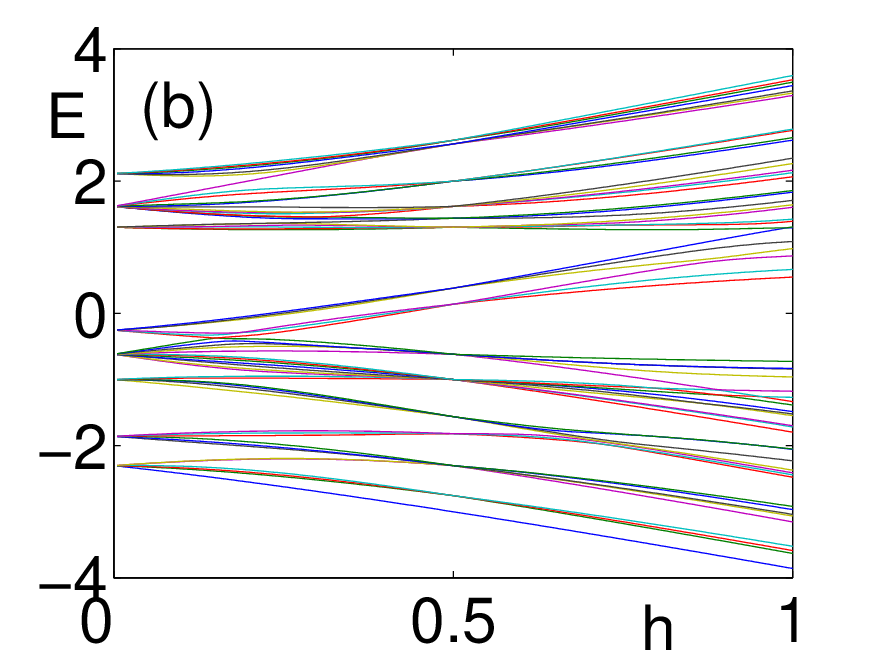}
\includegraphics{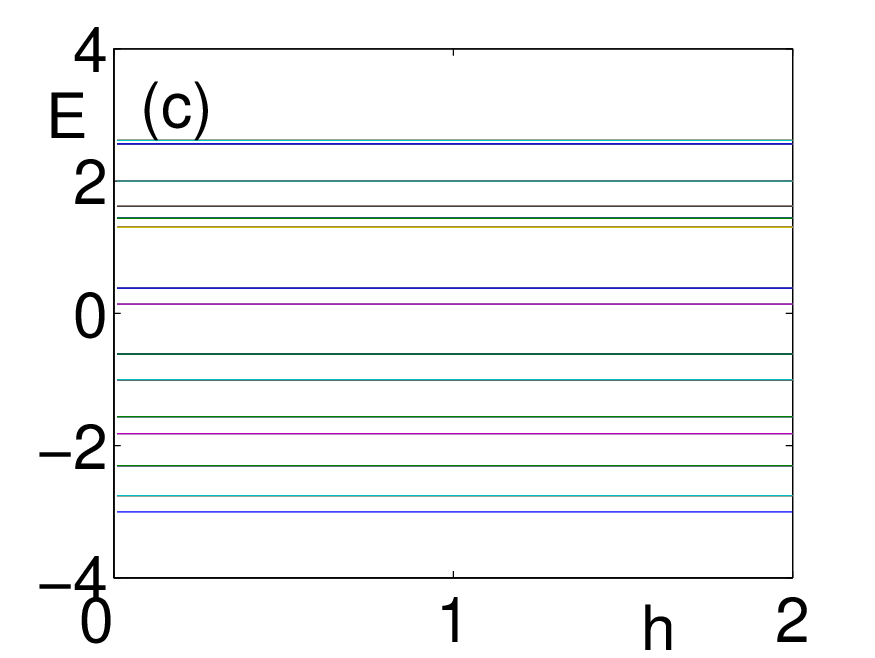}
 }
\caption{\label{C60} The spectrum of the C60 molecule as a function of the
double bond coupling, $h$, with single bond coupling equal to $1$. There are
six modes (two triplets) that are near zero energy. (a) With Kekul\'e
distortion. (b) With cuts between pairs of vortices. (c) With enlarged
vortices, where all double bonds are removed.}
\end{figure}
\end{center}

\begin{center}
\begin{figure}[h]
\resizebox{14cm}{!}
{\hspace{8cm}
\includegraphics{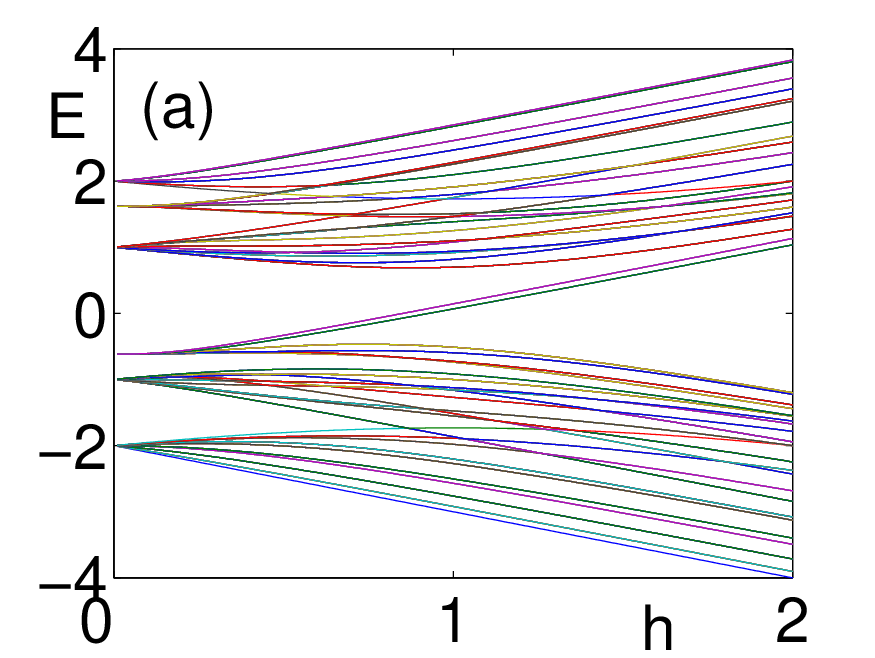}
\includegraphics{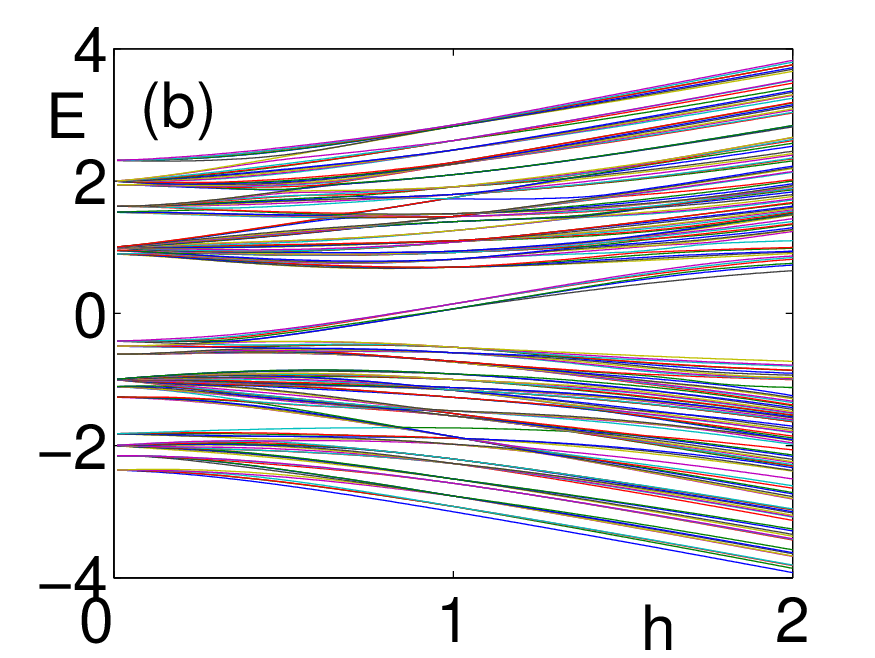}
\includegraphics{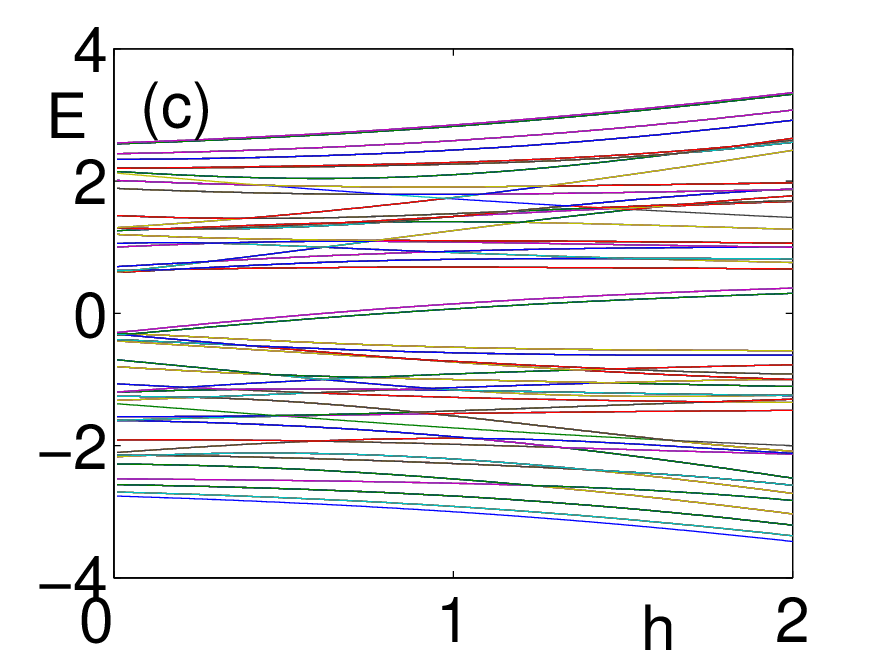}
 }
\caption{\label{C180} The spectrum of the  C180 molecule.
(a) With Kekul\'e distortion. (b) With cuts between pairs of vortices. (c)
With enlarged vortices.}
\end{figure}
\end{center}

In Fig.~\ref{C60Kekule}(a) we see the C60 fullerene with the Kekul\'e
distortion depicted in terms of single and double bonds. The vortices are
placed at the center of the pentagon plaquettes. In Fig.~\ref{C60Kekule}(b)
branch cuts are introduced between connecting pairs of vortices. In
Fig.~\ref{C60Kekule}(c) we increased the size of the vortices by removing
the double bonds immediately connected to the pentagons. Fig.~\ref{C60}
shows the energy levels of the C60 fullerene as a function of the strength
of the Kekul\'e distortion for the coupling configurations depicted in
Fig.~\ref{C60Kekule}. In Fig.~\ref{C180} the corresponding plots are
presented for the C180 molecule. In both plots we see that two lines
corresponding to two triplets have energies close to zero. These low lying
states are interpreted as zero modes in the continuous limit. We clearly see
that the insertion of branch cuts and the increasing of the size of the
vortices makes the low lying energy states to be even closer to the zero
energy. The difference of their energy from being exactly zero is due to the
small size of the system. It is expected to converge to zero when larger
molecules are employed~\cite{Gonzalez}.

To summarize, we showed that an effective scalar field is induced in the low
energy description of graphene originating from a distortion in its
tunnelling couplings. Importantly, the effective gauge field that originates
from the insertion of pentagons and heptagons in the lattice couples
directly to the scalar field imprinting vortices. As a result, the
fullerene-type molecules have six half vortices. By employing the index
theorem we demonstrated that there should be one zero mode for each pair of
such vortices. While the effective gauge field is responsible for the
presence of the zero modes the scalar field assures that the corresponding
degenerate states are not locally distinguishable and remain localized
around the vortex.

\section{Charge fractionalization}

The topological interplay between gauge and scalar fields in the
relativistic description of graphene has, yet, another surprising
consequence. It leads to the fractionalization of charge to the value $e/2$
for each $2\pi$ vortex. Thereby, it gives a natural setting, where
topological effects, such as anyons, can emerge. The charge
fractionalization is not an actual mechanical breaking of the electrons. It
is rather a collective effect, appearing in the quantum state of the system,
which conspires to assign half a charge to each vortex. In the case of
finite systems fractionalized charges, $e/2$, should come in pairs as the
total charge should always be an integer. This is the case as, for a compact
system, $2\pi$ vortices came always in pairs. The fractionalization of
charge is defined with respect to the difference between the electron
density with and without the vortex. Nevertheless, it is not an
uninteresting statistical effect that appears, e.g. due to averaging the
wave functions that correspond to the presence and the absence of an
electron. What actually happens is that the wave function of the system has
fractional charge as an eigenvalue.

In this Section we present how the coupling between electrons and a vortex,
as described in the previous section for graphene, gives rise to charge
$1/2$. For that we employ the quantum field theory
description~\cite{JackiwTopo}. Consider the Dirac operator,
$H(\Phi_\text{v})$, with a scalar field that has a full vortex. The complete
set of orthonormal eigenstates of the Hamiltonian is then given by
$\Psi^\text{v}_E$ parameterized by the energy eigenvalue $E$. The index
theorem states that the presence of vortices gives rise to zero modes. We
assume that our system with a {\em full} vortex has only a {\em single} zero
mode, $\Psi_0$, related to it. We further assume that the eigenvalues $E$
are symmetric around zero, i.e. the system possesses particle-hole symmetry.
We also encountered this condition in Eqn. ({\ref{symmetry}) of the index
theorem.

In quantum field theory terms the state of the system is described by the
quantum field operator, $\hat\Psi$. For the case of a vortex this operator
can be expanded in the eigenmodes of the corresponding Dirac operator,
$\Psi^\text{v}_E$, in the following way
\begin{equation}
\hat\Psi^\text{v} = \sum_E(b_E\Psi^\text{v}_E+d^\dagger _E\Psi^\text{v}_{-E})+a\Psi_0.
\label{qstate}
\end{equation}
Due to the symmetry of the non-zero energy states, the modes
$\Psi^\text{v}_{E}$ and $\Psi^\text{v}_{-E}$ can be paired. The first one
appears with the particle annihilation operator, $b_E$, that acts on the
conducting band and the second appears with the anti-particle creation
operator, $d^\dagger_E$, that acts on the valence band. However, the zero
mode does not have a partner. It is represented simply by the fermionic
operator $a$. The zero mode, as all the fermionic modes, can be either
occupied or empty. Since both states have zero energy, the zero mode has a
double degeneracy. Denoting the empty zero mode state by $|-\rangle$ and the
filled one by $|+\rangle$ we can represent the creation, $a^\dagger$, and
annihilation, $a$, operators as
\begin{equation}
a|+\rangle=|-\rangle,\,\,a^\dagger|+\rangle=0,
\,\,a|-\rangle=0,\,\,a^\dagger|-\rangle=|+\rangle.
\end{equation}

In order to determine the total charge of the system we employ the charge
operator, $Q$. This operator is given by
\begin{equation}
Q = {1\over2}\int d^2x (\hat\Psi^\dagger \hat\Psi-\hat\Psi\hat\Psi^\dagger),
\end{equation}
where the Schwinger's prescription is employed for its normalization. For
the state~(\ref{qstate}) the charge operator becomes
\begin{equation}
Q^\text{v} = \sum_E(b^\dagger_E b_E-d^\dagger_Ed_E)+a^\dagger a-{1\over 2}.
\end{equation}
Due to the balance between the energies in the conducting and valence bands,
their contributions to the charge of half filled systems cancel out. Thus,
the eigenvalues of the charge operator depend only on the state of the zero
mode, giving
\begin{equation}
Q^\text{v}|-\rangle= -{1\over 2}|-\rangle,
\,\,Q^\text{v}|+\rangle={1\over2}|+\rangle.
\end{equation}
This demonstrates that the vortex charge is $1/2$ in units of the electron
charge! Note that in the absence of the zero mode the neutrality condition,
$Q=0$ would arise and no fractionalization would take place.

The above result demonstrates that the fractional value of the charge is
indeed an eigenvalue of the charge operator. Our derivation was simply based
on the spectral symmetry of the Dirac operator and the presence of the zero
mode in the vortex sector, which is topological in its origin. Hence, the
result is completely general regardless of the shape of the vortex or other
details of the system. Fractional quanta of charge have been experimentally
measured in the case of fractional quantum Hall effect~\cite{Tsui,Goldman}.

\section{Anyonic statistics}

The fractionalization of charge in two dimensions gives rise to another
exotic effect. It makes it possible to realize anyons. The latter are
quasiparticles, such as vortices, with fractional statistics that can range
continuously from bosonic to fermionic. This is manifested by a non-trivial
phase factor acquired by the state of two anyons when one is circulating the
other, i.e. after two successive exchanges, as seen in Fig.~\ref{fig:anyon}.
In the case of bosons or fermions this phase factor would be 1 either
arising from bosonic, $(+1)^2=1$, or from fermionic, $(-1)^2=1$, statistics.
\begin{figure}[h]
\begin{center}
\includegraphics[width=8cm]{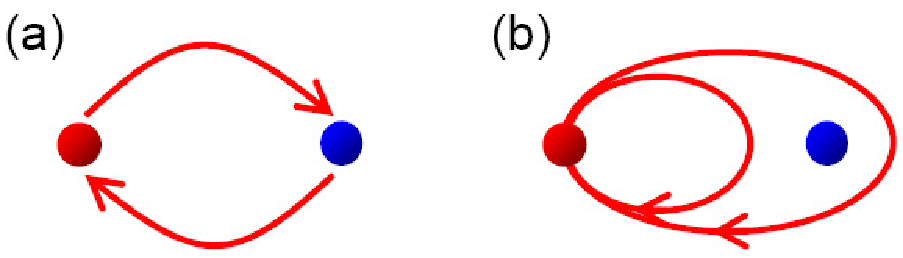}
\caption{(a) Two quasiparticles are clockwise exchanged. The wave function
of the quasiparticles is appropriately evolved according to their mutual
statistics. (b) A succession of two exchanges gives rise to a single
circulation of one particle around the other. In three spatial dimensions
this circulation can be continuously transformed into a smaller loop and
finally to a trivial one, where no circulation takes place. The latter
evolution is equivalent to no transportation at all so the evolution
operator should be the identity. Hence, the only possible exchange
statistics is the bosonic, $(+1)^2=1$, or the fermionic, $(-1)^2=1$, as they
are the only ones that give a trivial evolution after two exchanges.}
\label{fig:anyon}
\end{center}
\end{figure}
In the case of anyons, the phase resulting from two successive exchanges can
actually take any complex value. This possibility arises in two dimensions.
There, one can think of anyons as composite objects of a magnetic flux and
an electric charge. Then the phase factor acquired by the circulation can be
attributed simply to the Aharonov-Bohm effect. Indeed, the charge of one of
the anyons circulates the flux of the other one, as seen in
Fig.~\ref{fig:anyon1}. The acquired phase factor depends only on the number
of circulations and not on other characteristics of the path as expected
from a statistical evolution.
\begin{figure}[h]
\begin{center}
\includegraphics[width=10cm]{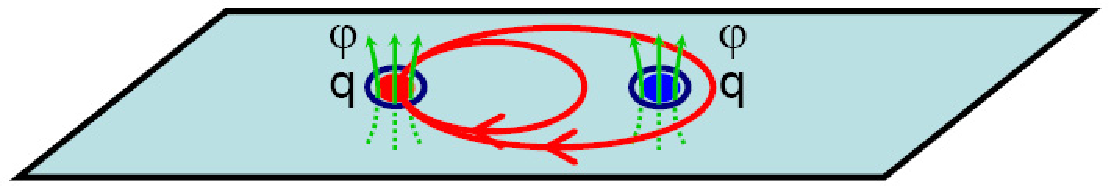}
\caption{In two dimension two exchanges lead to a
trajectory that cannot be continuously deformed to the identity one due to
the presence of the circulated particle. Thus, an arbitrary phase factor,
$e^{i\theta}$ could be assigned to this evolution giving rise to anyonic
statistics. Intuitively, this can arise from attaching an effective integer
flux, $\varphi$, to each anyon and an effective fractional charge $q$. Then
the Aharonov-Bohm effect is responsible for the phase factor $e^{i\theta}$
acquired by the wave function.}
\label{fig:anyon1}
\end{center}
\end{figure}

Consider the case of graphene with Kekul\'e distortion in the tunnelling
couplings. This gives rise to a scalar field that can support vortices.
Suppose that vortices can appear with vorticity $2\pi$. Then they will be
accompanied by half a charge as we saw in the previous Section. Circulating
such vortex around an identical one will give a phase factor $\pi$ due to
the Aharonov-Bohm effect,
$$
q\int\!\!\!\! \int_{S} {\bf B}\cdot d{\bf S} = {1 \over 2} 2\pi = \pi.
$$
This is a non-trivial phase that clearly demonstrates the anyonic nature of
the vortices~\cite{Franz1}.

The generation and detection of anyons in graphene is an exciting subject of
current research~\cite{Chamon,Jackiw,Chamon1,Pachos3,Ghaemi,Franz}. If
successful, it will give rise to a rich and versatile platform to perform
experiments related to the fractional quantum Hall effect. Indeed, graphene
offers several advantages, such as high electron mobility and resilience to
temperature, that can prove valuable for probing these strongly correlated
effects. Moreover, certain models with anyonic properties are of much
interest in topological quantum computation. The latter promises to overcome
the problem of environmental decoherence in the most efficient way.

\section{Conclusions}

In contrast to graphite, which is a rather soft substance, its constituent
graphene layers have a robust lattice structure. This gives rise to
graphene-based molecules  that have a variety of stable geometrical
configurations, such as the fullerenes and the nanotubes. The predominant
honeycomb lattice configuration of graphene is responsible for the effective
relativistic behavior of its electrons. This behavior, combined with the
long range quantum coherences present in graphene, allowed us to probe
several exotic phenomena. Indeed, graphene and its geometrical variants
provide an exciting laboratory for realizing numerous topological effects.

Initially, we saw how the low energy of graphene can be described by the
relativistic Dirac operator. An effective magnetic field naturally appears
whenever we introduce a geometrical deformation on the honeycomb lattice.
This magnetic field is naturally coupled to the Dirac fermions allowing the
emergence of phenomena already known from the quantum Hall effect. By
considering the underlined lattice we were able to relate local geometrical
characteristics of graphene to the general topology of the molecule by
employing the Euler character. As a result, the celebrated index theorem
provides insight into the spectrum of graphene molecules just by considering
their topology. This theorem is able to predict, in a simple way, the number
of zero modes that are present in the geometrical variants of graphene. We
compared these results with the spectra of fullerenes and nanotubes and
found remarkable agreement. The spectra of more complicated molecules can be
also predicted with our formalism.

Subsequently, we considered the effect of tunnelling distortions on the
relativistic properties of graphene. In particular, a scalar field emerges,
on which vortices can be imprinted. An effective magnetic field due to
distortions of the lattice geometry can create non-trivial vorticity on the
scalar field. We analyzed this effect for the case of fullerenes, where
tunnelling distortions are expected to appear due to the simultaneous
presence of pentagonal and hexagonal plaquettes. An appropriately modified
index theorem can be employed that correctly predicts the number of zero
modes. The number of zero modes is unchanged when the tunnelling distortion
is introduced. However, the wave functions of the zero modes are expected to
become localized around vortices. This makes the exotic properties that
accompany zero modes, such as the charge fractionalization, to become
properties of the corresponding vortices.

The generation of vortices can also appear in flat graphene sheets. There
vortices with $2\pi$ vorticity can be imprinted with the help of external
magnetic fields. The resulting system behaves similarly to the fractional
quantum Hall effect. The latter comprises of a continuum of two dimensional
electrons at very low temperatures that are subject to a vertical magnetic
field. There, vortices can support a variety of statistical behaviors. In
the case of graphene we saw that the zero modes localized to vortices are
responsible for the charge fractionalization as well as the emergent anyonic
statistics of the vortices. It is an exciting perspective to realize such
statistical phenomena with graphene. Its versatile behavior could prove
valuable to technological applications of these many body effects.

\section*{Acknowledgements}

This work was supported by the EU grants EMALI and SCALA, EPSRC and the
Royal Society.

\medskip

\end{document}